\documentclass[final,5p,times,twocolumn]{elsarticle}  

\usepackage{color,graphicx}

\usepackage{amssymb}
 \usepackage{amsthm}

\newcommand\etal{{\it et\ al.\ }}

\journal{..... }

\begin{document}

\begin{frontmatter}


\title{ 
Developing ZnO Nanoparticle embedded Antimicrobial Starch Biofilm for Food Packaging}
\author[uicet]{Prakash Kumar}
\author[uicet]{Sanjeev Gautam}\ead{sgautam@pu.ac.in}
\address[uicet]{Advanced Functional Materials Lab., Dr. S.S. Bhatnagar University Institute of Chemical Engineering \& Technology, Panjab University, Chandigarh 160-014, India}
\cortext[cor1]{Corresponding author: Phone/Fax: ++91-9779713212, }

\begin{abstract}
Starch based bio-films act as biodegradable packaging material for coating and wrapping food products, which is alternate to the synthetic polymer. Nanoparticles (NPs) are also alternative to antibiotics to target pathogens. Through interaction of nanoparticles with food packaging material the overall quality of the packaged food also get enhanced. In this research work, ZnO NP embedded corn starch films were developed. First, ZnO NP were synthesized by two approaches (a) Sol-gel approach and (b) Green synthesis approach; and then ZnO NP embedded corn-starch bio-films were developed. In sol-gel technique ZnO NP were synthesized at different pH ranging from 8 to 11, with an average crystallite size of 28 to 53 nm. In green synthesis technique ZnO NP were synthesized by using Azadirachta indica (Neem) at different concentration of 2\% and 3\% with an average crystallite size of 36 nm and 32 nm, respectively. Characterization of synthesized ZnO NPs were done with XRD, UV-DRS, UV-VIS, FTIR and FE-SEM. The functional properties of the ZnO embedded starch film were enhanced with addition of 5\% citric acid solution(w/w) as cross-linker and 0.90 g glycerol(w/w) as plasticizer. Film-F1 embedded with chemically synthesized ZnO NP, Film-F2 embedded with biochemcially synthesized ZnO NP and Film-F3 without being embedded by any NPs. Characterization of synthesized films were done with SEM and XRD. Film solution of F2 showed higher antimicrobial effectiveness than F1; against E.coli and S.aureus bacterial strains with an inhibition zone of 14 mm and 6 mm. 7days biodegradability analysis of the films were also done. Along with this, their current application and future perspectives in the food sector are also explored.

\end{abstract}

\begin{keyword}
Starch \sep Biofilms \sep ZnO Nanoparticles \sep Chemical synthesis \sep Green synthesis \sep ROS \sep Sol-gel \sep \emph{Azadirachta indica} \sep Functional properties \sep 9-point hedonic scale


\end{keyword}

\end{frontmatter}


\section{Introduction}
Starch based food packaging material with embedded NPs (Nanoparticles) attracted the attention of all researchers across the world due to its antibacterial and biodegradable properties. Starch based food packaging materials are alternates to the synthetic polymers in food packaging application revolution. Petroleum based synthetic polymer are non-biodegradable in nature and its wide application as food packaging material increases the attention of the researchers towards environment waste disposal crisis \cite{r1}. Agro based by-products drawn the researcher's attention towards synthesis of the biomass based food packaging materials which are alternative to synthetic polymer. Biomass based food packaging material are cheap, easily available, eco-friendly and abundant in nature. Biodegradable starch films, as a food packaging materials are eatable due the presence of starch and other plants based by-products like lipids, proteins and polysaccharides \cite{r2}. Many researchers are developing bio composites from renewable resources, producing more sustainable and environmental friendly materials has gained the attention at the international level. Synthetic based food packaging materials as solid waste creates severe environmental problem due to its non-biodegradability property. Plastic packaging is one of the major causes of sea pollution as over 67 million tons of polymeric packaging waste has resulted in environmental issues \cite{r3}. Globally, there are special rules and regulations to dispose synthetic polymeric waste (like plastic) but less than 5\% of the whole get recycled \cite{r4}.

Therefore, biopolymer such as starch, cellulose, chitosan and sodium alginate  and proteins such as gelatin, whey protein and soy proteins based food packaging system have been introduced since 1970 \cite{r5}. They are ample, cheap, and biodegradable, and isolated from living organisms such as plants and animals \cite{r6,r7,r8}. The consideration of people towards the biodegradable material can reduce the environmental pressure. It includes a search for alternatives to petroleum-based materials. The development of biomaterials holds great promise to mitigate many of the sustainability problems, offering the potential of renewability, biodegradation, and a path away from harmful additives. Packaging films and coatings from bio-polymers such as polysaccharides, proteins and fats had significantly growing to enhance the shelf life, overall quality and stability of the packaged food. Among polysaccharides starch has been studied as substitute Biomaterial for packaging due to its polymeric properties \cite{r9}. Hence, the demand of the bio-based polymer materials is increasing rapidly in order to solve the waste disposal issues to the controlled magnitude \cite{r10}.

The main function of the food packaging material is to protect the packaged food form the environmental factors such as temperature (high or low), pressure (high or low), light (high or low intensity), moisture (high or low), humidity (high or low), microorganisms, enzymes, insects or any other foreign materials \cite{r11}. Historically, the main mission of food packaging material is to keep the packaged food material safe, consumable and away from any physical, chemical and biochemical contamination. The microbial contamination of packaged food not only degrades the quality of food but also degrade the health of the consumer \cite{r12}. Recent developments in food packaging material not only improved the overall quality of the food packaging material but also enhance the shelf life of packaged food. World War I and World War II play an important role in improving the food packaging system \cite{r11}.

Active packaging materials like Oxygen scavengers (for packaging bakery products like breads, cake, cookies, biscuits etc), CO$_2$ scavengers (for packaging meat industries, snacks, coffee etc), Ethylene scavengers (for packaging fruits and vegetables), Ethanol emitter (for packaging pizza crust and most bakery products), Moisture absorbers (for packaging meat and meat products, snacks food, cereals, sandwiches, fruits and vegetables) and nanocomposites packaging material, all plays dynamic role in food preservation system \cite{r11,r12}. In recent years researchers have seen that nanotechnology playing an important role in filling the gap of innovative development in food packaging material \cite{r13}. Nanocomposites films are composed of NPs (of metals or non-metals) \cite{r14}.

In order to face these problems, renewable natural resources based biodegradable polymers were promoted as an attractive substitute for synthetic polymers. Therefore, biopolymer such as starch cellulose, chitosan, and sodium alginate etc and proteins such as gelatin, whey protein and soy proteins based food packaging system have been introduced since 1970 \cite{r5}. They are abundant, cheap, easily degradable in nature and isolated from living organisms such as plants and animals-environment issues due to its biochemical features to degrade.

Due to the reduction of the petroleum sources, alternative natural and renewable resources. On industries scales, different types of starch exist in nature such as potato, maize, rice, legumes starch have been started to produce which are commercially used in different industries such as food, pharmaceutical, and paper. Among all different variety of starch, Moth starch may be assumed as one of the most important source starch which is not fully explored. Moth beans are grown in arid and semi-arid regions in India such as Rajasthan, Haryana, Madhya Pradesh and also some region of Punjab and Uttar Pradesh \cite{r19} during the summer. Moth beans starch may be believed as one of the abandon sources of starch which is recently unheeded and usually utilized moth beans in cooking in India.

\begin{figure}[htb!]\centering
  \includegraphics[width=0.48\textwidth]{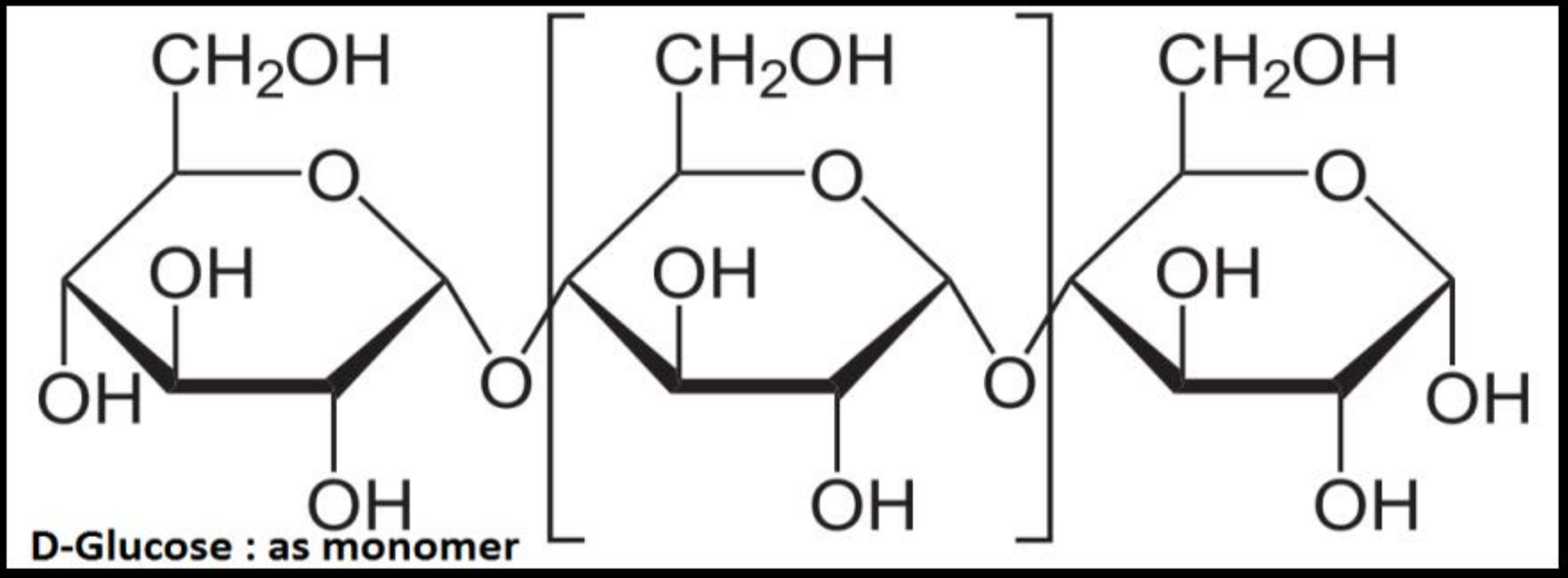}
  \caption{Sturcture of starch molecule which contain D-Glucose as a repeating monomer.}\label{fig4.1}
\end{figure}

Starch (as shown in Fig. \ref{fig4.1}) has acted a significant role in the human diet since before written history and consumed in papermaking and adhesive industries. In present time, starch used to produce new products such as biodegradable plastics, drug carrier and water absorbent polymer using emerging technologies. These growths may be made a new market for starch-based products. New starch-based products are biodegradable and eco-friendly than many synthetic polymer-based materials. Starch films are brittle so different type of plasticizer such as sorbitol, ethylene glycol and glycerol are added in the polymers to increase the elasticity. The plasticizers molecules entered in-between the polymer chains. Now, polymer chains were easily slide over another. Freely mobile structure caused the natural films more flexible \cite{r20}. In food packaging industries, utilization of starch in food packaging and coating are justified due to its nontoxic nature, fragrance-free, colorless and excellent film forming capability. Moreover, the starch may be utilized as carriers of natural antioxidant and antibacterial active components to produce smart packaging materials. Starch showed many properties such as outstanding gas barrier and film forming property. So it is worn to produce eco-friendly food packaging materials that have capabilities to substitute petroleum-based synthetic food packaging materials in future. However, some disadvantages of starch-based packaging material such as hydrophilicity, brittle nature, and low mechanical strength have restricted starch application in the food industry especially in packaging and coating \cite{r21}. To deal with these problems, starch modification techniques are used to improve the functional properties such as Film thickness (FT), moisture content (MC), solubility (S), swelling index (SI), water vapor permeability (WVP), opacity (OP), Tensile Strength (ST) and biodegradability of the starch films.

Plasticizer is polyols materials such as sorbitol, ethylene glycol and glycerol that are added in the polymers to increase the elasticity. Since natural polymers based films and coatings are brittle because of larger number of hydroxyl groups in the matrix. The plasticizers molecules entered in-between the polymer chains. Thus polymer chains were interrupted resulting to improve the free space between the chains. Now, polymers chains were easily slide over another. Freely mobile structure caused the natural films more flexible \cite{r20}.
\section{Materials and Methods}
\subsection{Chemical Synthesis of ZnO NPs through Sol-Gel Approach}
\begin{table*}[htb!]\centering
\caption{Chemicals used in the synthesis of the ZnO NPs in sol-gel approach.} \vskip 0.5cm
\begin{tabular}{|c|c|c|c|c|} \hline
S.No.   &Chemical Name         &Chemical formula      &Molarity&Materials taken \\
        &                      &                       &          &(/100 ml of DI H$_2$O)  \\ \hline
1       &Zinc acetate dihydrate &Zn(CH$_3$OO)$_2\cdot$2H$_2$O &1M&21.951 g                  \\ \hline
2.	    &Potassium hydroxide    &KOH                  &1M&05.611 g                  \\ \hline
3.	    &Ethanol (for washing)  &C$_2$H$_5$OH          &-& -                         \\ \hline
\end{tabular} \label{tab3.1}
\end{table*}
\begin{itemize}
\item {\textbf{Methodology:}}
The precursor used for the sol-gel synthesis of ZnO nanoparticles is zinc acetate dehydrate with 99.9\% purity. 1 Molar solution of zinc acetate dihydrate solution is prepared by dissolving 21.951 g in 100 ml of deionized water under continues magnetic stirring for 15 minutes. 1 Molar solution of potassium hydroxide solution  is prepared by dissolving 5.611 g  in 100 ml of deionized water under continues magnetic stirring for 15 minutes. Titration is done to get the precipitation reaction between zinc acetate dehydrate solution and potassium hydroxide solution until we get the desired pH (like 7, 8, 9, 10 and 11) with continues stirring over hot plate magnetic stirrer at 30$^\circ$C. After getting desired pH, titration process is stopped and then putted on continues stirring for 2-3 hrs and then settling is done for 24 hrs. Centrifuge is done at 5000 rpm for 3 minutes and washed several time (nearly 3 to 4 times) with absolute ethanol to remove the water soluble impurities present inside the sol. Heating or drying of sol at 80$^\circ$C for 15 hrs is done in hot air oven to get ZnO NPs. After heating, water vapors from Zn(OH)$_2$ get evaporated to get ZnO, after that  crushing is done over mortar and pestle. At the end annealing is done at 450$^\circ$C for 1hrs in a muffle furnace to get ZnO NPs. Table \ref{tab3.1} gives the chemicals used in the synthesis of the ZnO NPs in sol-gel approach.
The flow sheet of the synthesis ZnO NPs by sol-gel process is described below in Figure \ref{fig3.1}.
\end{itemize}
\begin{figure}[htb!]\centering
  \includegraphics[width=0.48\textwidth]{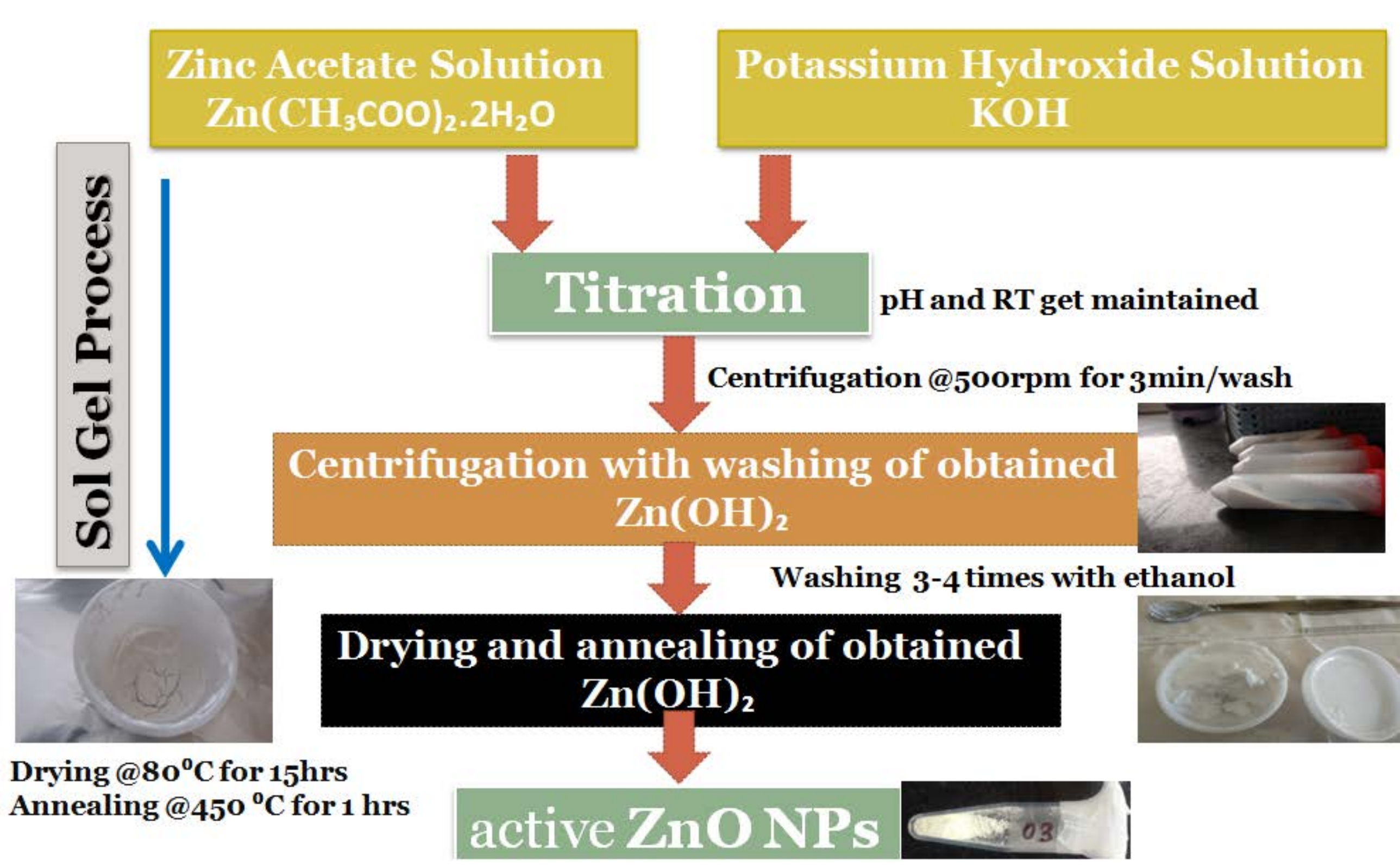}
  \caption{Synthesis of ZnO NPs by chemical process (sol-gel process).}\label{fig3.1}
\end{figure}
\subsection{Biochemical Synthesis of ZnO Nanoparticles through Green Synthesis Approach}
\begin{table*}[htb!]\centering
\caption{Chemicals used in the synthesis of the ZnO NPs in Green synthesis approach.} \vskip 0.5cm
\begin{tabular}{|c|c|c|c|c|} \hline
S.No.   &Chemical Name                   &Chemical formula     &Molarity&Materials taken          \\
        &                                &                     &         &(/100ml of DI H$_2$O)\\ \hline
1       &Zinc acetate dihydrate          &Zn(CH$_3$OO)$_2\cdot2H_2$O&2mM     &0.043902 g               \\ \hline
2.	    &Potassium hydroxide             &KOH                  &1M&5.611 g                 \\ \hline
3.	    &Ethanol (for washing)           &C$_2$H$_5$OH         & -       & -                        \\ \hline
4.      &\emph{Azadiarchta indica} leaves& -                   & -       &2 g and 3 g              \\
        & (Neem leaves)                  &                     &         &                          \\ \hline
\end{tabular} \label{tab3.2}
\end{table*}
\begin{itemize}
\item {\textbf{Methodology:}}
In this method fresh \emph{Azadiarachta indica} leaves (Neem leaves) washed and then cut into small pieces. The fine pieces crushed or grinded with help of mortar pester.  After that 2 g and 3 g of that paste was dissolved into the 100 ml of deionized water separately to get the 2\% and 3\% concentrate solution. Boiling was done at 100 $^\circ$C for 10 minutes to get the pure neem extract. Then filtration was done with 1whatman filter paper 2-3 times to separate the desired pure neem extract form the concentrate solution.
\begin{figure}[htb!]\centering
  \includegraphics[width=0.48\textwidth]{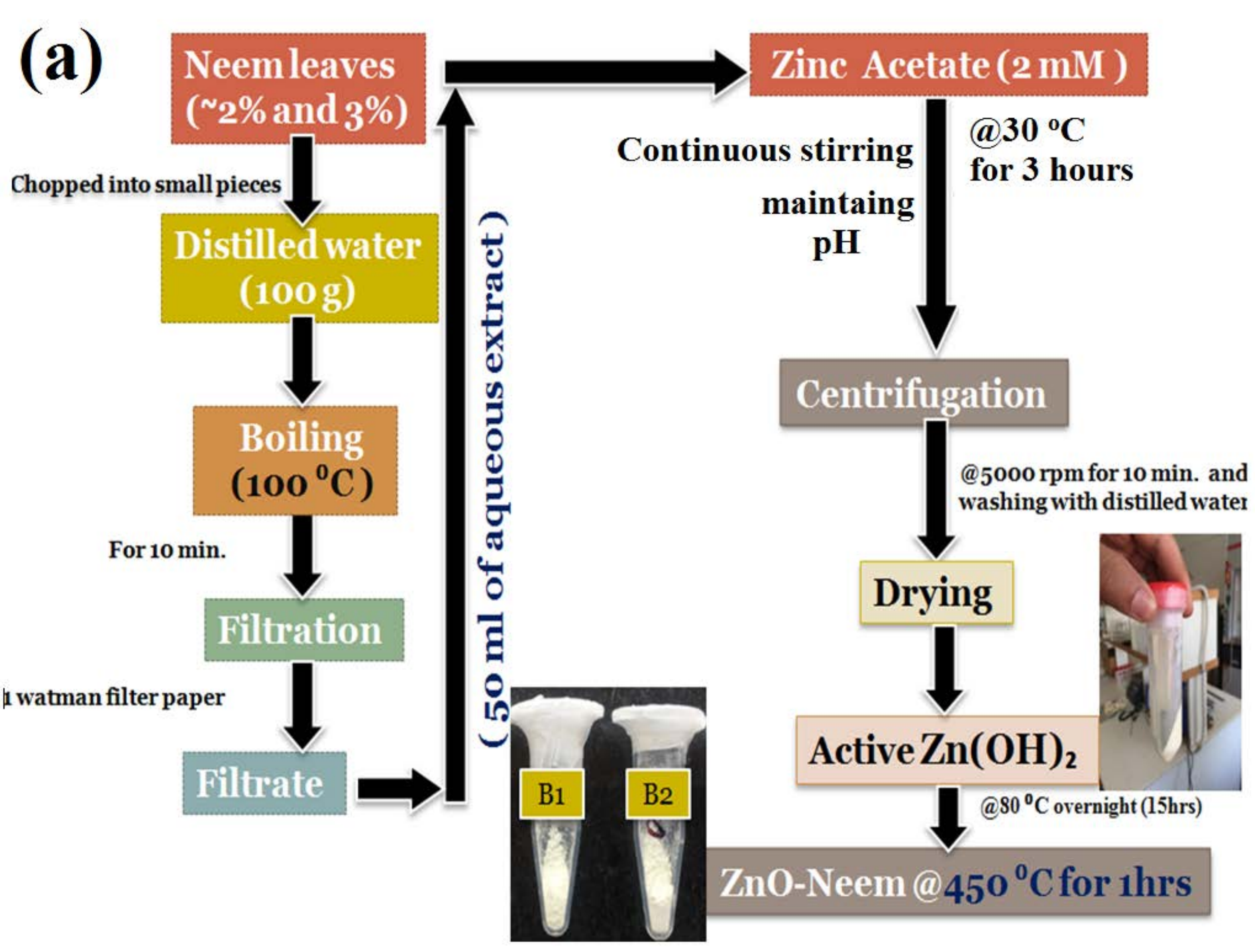} \\
  \includegraphics[width=0.48\textwidth]{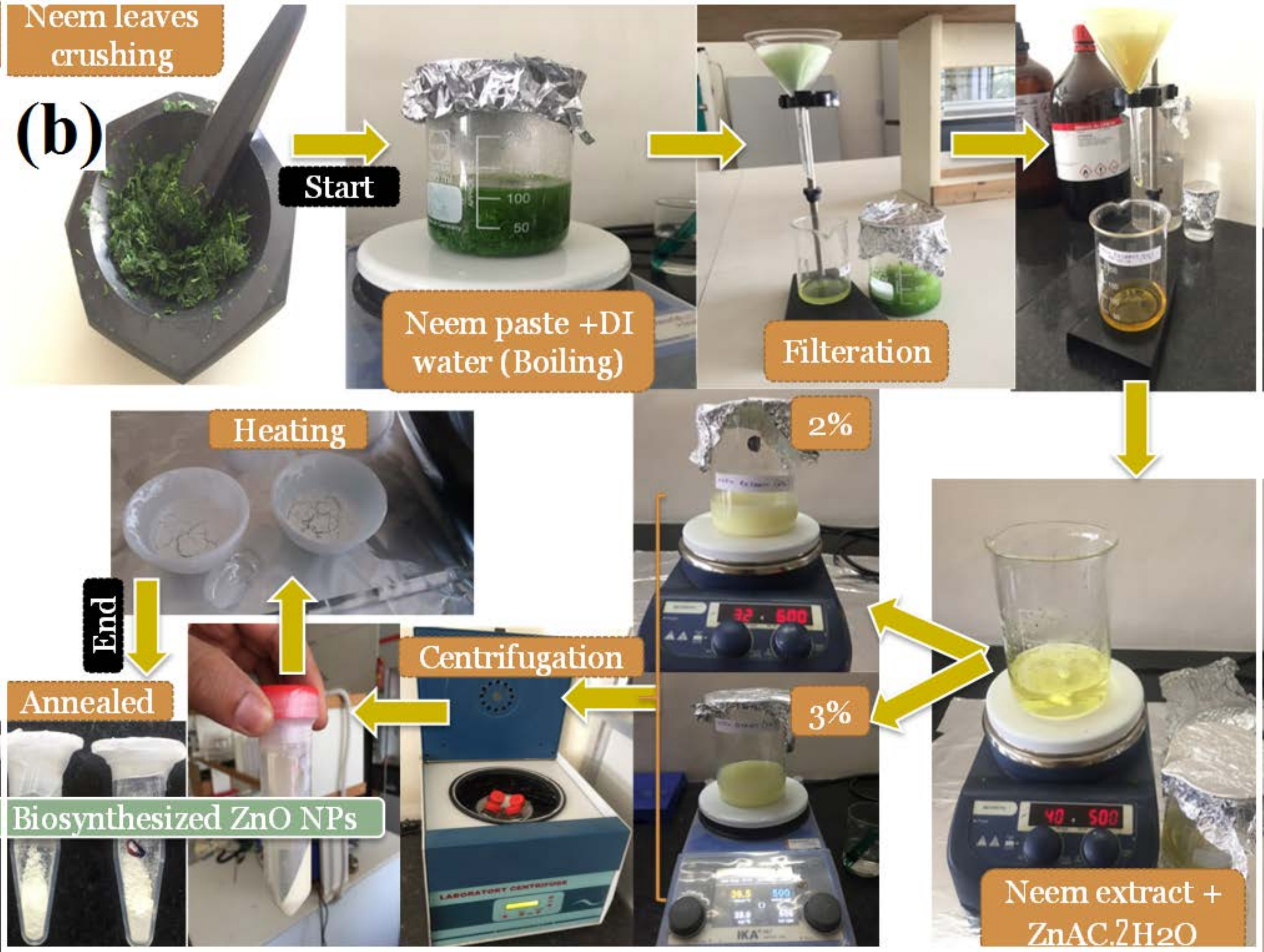}
  \caption{(a) Flow sheet of synthesis of ZnO NPs with the help of \emph{Azadiarachta indica} (Neem);
 (b) complete green synthesis process followed in AFM Lab, Dr. S.S.B. UICET, PU, Chandigarh, India; for the synthesis of ZnO NPs.}\label{fig3.2}
\end{figure}
Then this 50 ml of filtrate was taken as medium to prepare 2 mM solution of the Zinc acetate dihydrate solution was made with continues stirring over the hot plate magnetic stirrer at 30$^\circ$C. Then pH was maintained by adding 1M potassium hydroxide solution drop wise into the zinc acetate dihydrate-neem extract solution. Precipitate formation is seen in basic media ($\sim$ 10 pH to 11 pH). Centrifuge is done at 5000 rpm for 3 minutes and washed several time (nearly 3 to 4 times) with absolute ethanol to remove the water soluble impurities present inside the sol. Heating or drying of sol at 80$^\circ$C for 15 hrs is done in hot air oven to get ZnO NPs. After heating, water vapors from Zn(OH)$_2$ get evaporated to get ZnO, after that  crushing is done over mortar and pestle. At the end annealing is done at 450 $^\circ$C for 1hrs in a muffle furnace to get ZnO NPs. Flow sheet of the process is shown in Fig. \ref{fig3.2}(a). Tab. \ref{tab3.2} gives the chemicals used in the synthesis of the ZnO NPs in Green synthesis approach.
\end{itemize}

\subsection{Development of ZnO NPs Embedded Food Packaging Material (FPM)}
\begin{table*}[htb!]\centering
\caption{Material taken in synthesis of ZnO-corn starch film.} \vskip 0.5cm
\begin{tabular}{|c|c|c|c|} \hline
S.No.   &  Chemical Name & Chemical formula         &Materials taken \\
        &                &                          &(per 100ml of D.I. water) \\ \hline
1       & Corn starch    & [(C$_6$H$_{10}$O$_5$)$_n$] & 3 g                       \\ \hline
2.	    & Citric acid    & (C$_6$H$_8$O$_7$)        & 0.15 g                    \\ \hline
3.	    & Glycerol       & (C$_3$H$_8$O$_3$)        & 0.90 g                    \\ \hline
4.      & Zinc oxide NPs & ZnO                      & 0.100 g or 100 mg         \\ \hline
\end{tabular} \label{tab3.3}
\end{table*}
\begin{itemize}
\item {\textbf{Methodology:}}
The principle method for preparing ZnO NPs embedded starch films was based on solution casting and evaporation method (as shown in Fig. \ref{fig3.3}). For this process, 3 g of corn starch (CS) was mixed with 100 ml deionized water under continuous stirring at 300 rpm for 15 minutes. Then prepared 100 mg of ZnO NPs is added to the solution after this immediately 0.90 g of Glycerol (of 30 wt\% of CS) and few drops of 5\% Acetic acid solution (to maintain pH 3 to 4 of the final solution) were mixed and heated at 90 $^\circ$C until the mixture was gelatinized for 30-45 minutes.
\begin{figure}[htb!]\centering
  \includegraphics[width=0.48\textwidth]{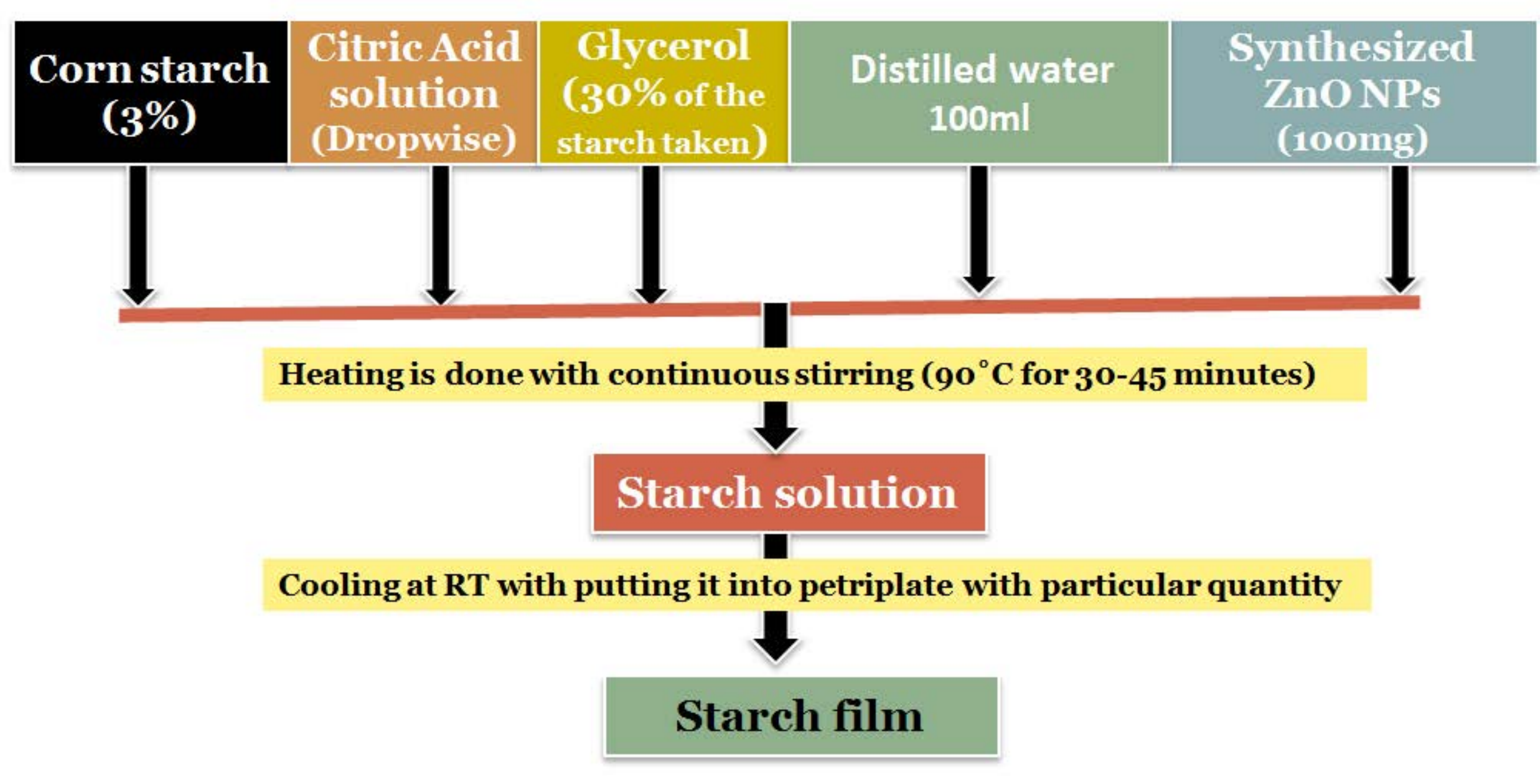}
  \caption{Process flow diagram of synthesis of ZnO embedded corn starch film.}\label{fig3.3}
\end{figure}
Then the prepared solution is sonicated immediately at 90 $^\circ$C for 5minute and then degassing of prepared solutions was done using vacuum oven. Lastly, prepared homogeneous solutions were poured and spread evenly onto a petri dish and dried in an incubator at 25$\pm$1$^\circ$C for 24 hours. Dried films were peeled off from Petri dishes and conditioned in a desiccator at 25$\pm$1$^\circ$C until further analysis (Li \etal, 2018; Kumar \etal, 2019). Material taken in synthesis of ZnO-corn starch film is given in Tab. \ref{tab3.3}.
\end{itemize}

\subsection{Analysis of Functional Properties of the Synthesized Films}
\begin{figure}[htb!]\centering
  \includegraphics[width=0.48\textwidth]{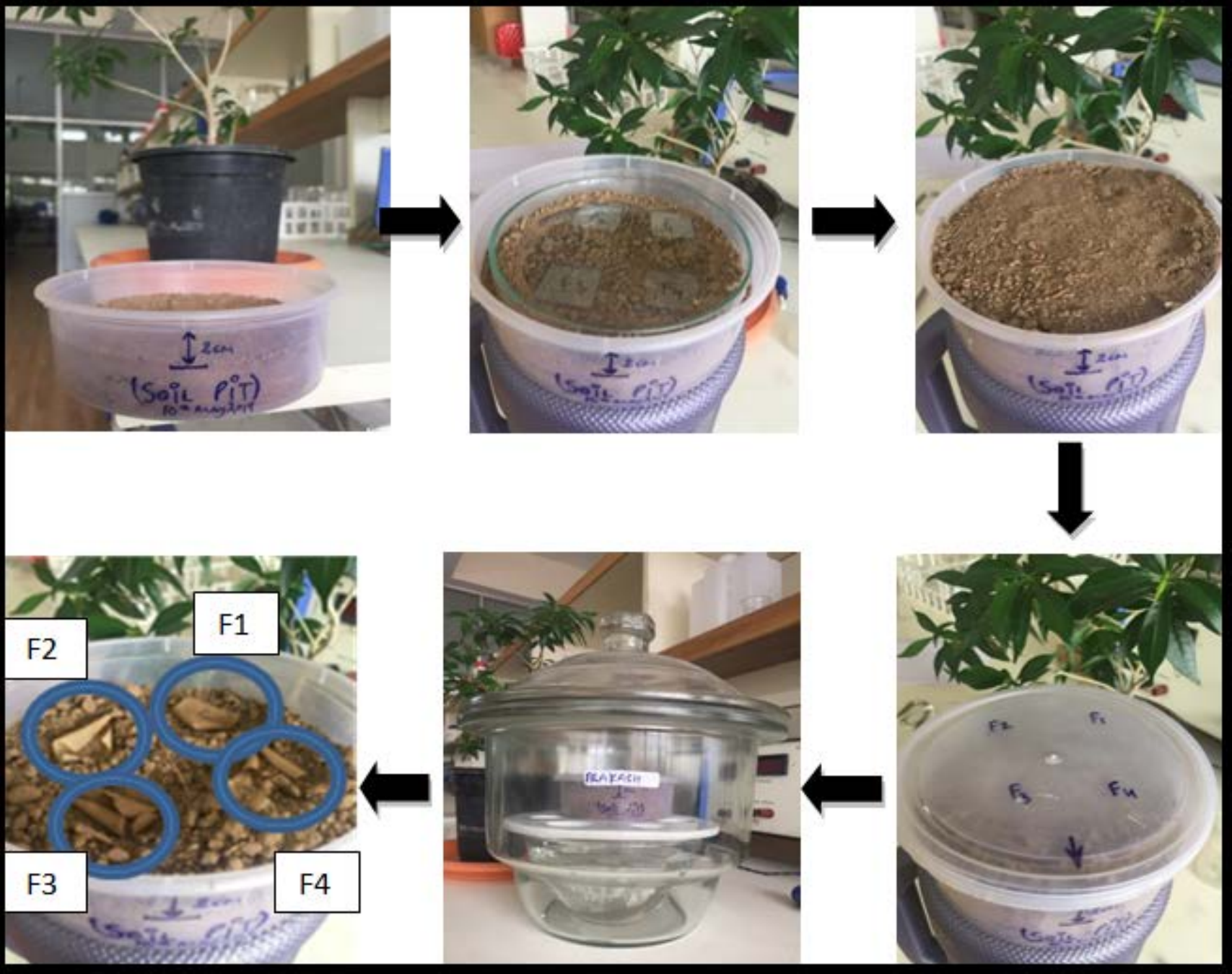}
  \caption{Biodegradation of the synthesized films after 7 days.}\label{fig3.4}
\end{figure}

\begin{enumerate}
\item {\textbf{Film thickness (FT):}}
Thickness was measured using a digital micrometer with an accuracy of $\pm$ 0.001 mm (Mitutoyo, Japan). The thickness of each samplewas evaluated at six different positions and then averaged of all thickness for analysis.

\item {\textbf{Moisture content (MC):}}
The moisture content (MC) of the films was measured in terms of weight loss. Briefly, 2*2 cm$^2$ samples were cut from each film and then weighed. Then, all samples were dried in a hot air oven at 105$\pm$1$^\circ$C for 24 hrs and weighed again. MC was calculated using equation (\ref{eq3.1}):\\
\begin{equation}
Moisture ~Content (MC) =(w_w/w_d-1)\times100 \label{eq3.1}
\end{equation}
Where w$_w$ is the mass of the pre-dried samples and w$_d$ is the mass of the dried sample (Chiumarelli and Hubinger, 2014) \cite{r179}.

\item {\textbf{Swelling index (SI):}}
Swelling index indicates the interaction between the starch and water molecules and evaluated using described method (Cao \etal, 2007) \cite{r130} with small modification. In brief, specimens with 2 cm*2 cm dimension were dried in hot air oven at 105$\pm$1$^\circ$C for 24 hrs and weighed. Dried specimens were immersed in distilled water for 2 min and then removed from the distilled water. An excess amount of water was removed from the swelled samples and weighed. The quantityof water absorbed by the specimens were evaluated according to equation (\ref{eq3.2}):\\
\begin{equation}
Swelling ~Index (SI) = [(m_2-m_1)/m_1]\times100 \label{eq3.2}
\end{equation}
Where, m$_2$, m$_1$is the weight of swelled samples after removal of excess water and weight of dried samples respectively, each measurement was done in triplicate.

\item {\textbf{Optical property (OP):}}
Film opacity was measured using the Hunter Lab colorimeter (Color Flex EZ, Hunter Lab, USA). The measurements were made after calibration of the equipment with a white and a black background. L*, a* and b* values of the films from Food Lab, color space were determined using a D65 illuminant and 10 $^\circ$ standard observers. The opacity of film was determined according to the following equation (\ref{eq3.3}):\\
\begin{equation}
Opacity~ (OP) = [O_{pb}/O_{pw}]\times100 \label{eq3.3}
\end{equation}
Where, O$_{pb}$ is the opacity of film against a white background; O$_{pw}$ is the opacity of the film against a black background (Fakhouri \etal in  2015 and Babaee \etal in 2015) \cite{r116,r119}.

\item {\textbf{Solubility (S):}}
Water solubility of the film was measured according to the described method by (Wang \etal, 2017) \cite{r168} with some modifications. In short, four samples of each film with 2*2 cm$^2$ dimension were cut from the films. All prepared samples were dried in a hot air oven at 105$\pm$1$^\circ$C for 24 hrs and weighed (w$_0$) with an accuracy of $\pm$ 0.0001. Dried samples were immersed in a beaker containing 15 mL of double distilled water and stored at 25$\pm$1$^\circ$C for 24 hrs. Then, swelled samples were removed and dried again at 105$\pm$1$^\circ$C for 24 hrs in a hot air oven and weighed (w$_1$) again. Aqueous solubility (S\%) was evaluated by equation (\ref{eq3.4}):\\
\begin{equation}
Solubility~ (S) = [(W_0-W_1)/W_0]\times100 \label{eq3.4}
\end{equation}
Where, W$_0$ is the weight of the dried sample before water immersion and W$_1$ is the dry weight of the insoluble sample after immersion \cite{r191}.

\item {\textbf{Water vapor permeability (WVP):}}
Water vapor permeability (WVP)was performed by adopting described method (Colivet and Carvalho, 2017) \cite{r171} with small changes. In short, specimens were separated from films and then placed on the mouth of modified glass beaker with vacuum grease containing dried calcium chloride (RH 0). Thereafter, Sample loaded beakers were placed in a desiccator containing distilled water (RH 100). Thus, desiccators were kept in an incubator (Innova-4230, New Brunswick, NJ, USA) at 25$\pm$1$^\circ$C temperature. The gain in weight of the beakers was recorded at a fixed interval of time until the constant weight of the beakers was attained. Water vapor barrier through the film was calculated by using equation (\ref{eq3.5}):\\
\begin{equation}
WVP = ((WVTR\times L)/(P_a-P_b )) \label{eq3.5}
\end{equation}
WVTR is water vapor transmission rate, L is the thickness of specimens and the P$_a$-P$_b$ difference in the pressure outside and inside the sample beakers, all measurements were performed in triplicates \cite{r191}.

\item {\textbf{Tensile Strength (TS):}}
Tensile strength and elongation-at-break (E \%) of the edible film was measured with the help of Texture Analyzer (TA.XT Plus, Stable Microsystems) with cell load (5 kg) and a crosshead speed of 0.5 mm/s. Tensile strength (TS) and elongation at break (E \%) of the films were calculated according to reported method (Farhan and Hani, 2017) \cite{r180}. In brief, films samples (1.5 cm wide x 7 cm long) were cut to film. All samples were put in a desiccator for conditioning at RH 70\%, 25$\pm$1$^\circ$C for 72 hrs. Conditioned film samples were mounted between the grips with an initial grips distance of 50 mm. Tensile strength and elongation were determined by applying following equations (\ref{eq3.6}) and (\ref{eq3.7}):\\
\begin{equation}
Tensile Strength(TS) = F/A  \label{eq3.6}
\end{equation}
\begin{equation}
Elongation (E) = [(X-X_1)/X_1]\times100  \label{eq3.7}
\end{equation}
Where, F is the maximum force at break point, A is the cross-sectional area of films (Thickness, Width), X$_1$ is the initial gaps in grips and X is the length of the film at breaking point. Three replicates were used to analyze for each formulation of the film \cite{r191}.

\item {\textbf{Biodegradability of film:}}
Biodegradability of samples was checked using soil buried test (as shown in Fig.\ref{fig3.4}). All specimens with dimension (1*1cm$^2$) of modified starch and composite were prepared from the films. Thereafter, plastic containers filled up to the surface with a mixture of uniform size soil and compost. Specimens were buried in the mixture below 2 cm from the mouth of the containers. Under controlled conditions (humidity 90$\pm$2\%, temperature 25$\pm$1$^\circ$C), weight loss of films were monitored after a fixed time interval of one week (Babaee \etal, 2015) \cite{r191}.

\end{enumerate}

\begin{table}[htb!]\centering
\caption{Different types of synthesized films.} \vskip 0.5cm
\begin{tabular}{|c|l|} \hline
Films &Film characteristics \\ \hline
F1    &Chemically synthesized ZnO NPs at pH 10 embed-\\
      &ded corn starch film in which smallest calculated \\
      &crystallite size of 28 nm of ZnO \\
      &NPs is taken (Ref. Table \ref{tab4.1}).   \\ \hline
F2	  &Biochemically synthesized ZnO NPs embedded- \\
      &corn-starch film (with B2: 32.68 nm crystallite-\\
      &size ZnO NPs at maintained pH 10). \\ \hline
F3	  &starch film without embedded any NPs.\\ \hline
F4    &Biochemically synthesized ZnO NPs embedded \\
      &corn-starch film (with B1: 36.34 nm crystallite \\
      &size ZnO NPs at maintained pH 10).\\ \hline
\end{tabular} \label{tab3.4}
\end{table}

\subsection{Characterization of ZnO NPs Embedded Starch Films}
\begin{enumerate}
\item {\textbf{X-Rray Diffraction (XRD) Analysis:}}
XRD pattern of starch was carried out using a XRD, X'Pert PRO, Panalytical, Netherlands with Cu K radiation at a voltage of 45 kV and 40 mA, and scanned in the range from 3 to 60$^\circ$ having scanning speed of 2$^\circ$/min.

\item {\textbf{Scanning Electron Microscopy (SEM) Analysis:}}
The microstructure of modified starch and composite films were quantified using SEM (JEOL JCM-6000 BENCHTOP, JEOL Ltd, Japan). Standard procedure was followed to perform the test with accelerating voltage of 10 kV. Images were captured in magnification range(300-2000 cm$^{-1}$) with a spot size of $~$2 nm.

\item {\textbf{Antimicrobial activity of synthesized ZnO NPs embedded starch films:}}
Antibacterial activity of ZnO embedded corn starch film against \emph{E. coli} (gram negative bacteria) and \emph{S. aureus} (gram positive bacteria) was studied using standard agar diffusion method. Antibacterial activity was performed against two bacterial strains as per guidelines (CLSI, M02-A12). These were grown overnight and diluted in Mueller-Hinton broth (MHB) to a cell density of 10$^5$ Colony Forming Unit (CFU)/mL. 100 $\mu$l of this culture was spread on the Mueller-Hinton Agar (MHA) plate and allowed to dry in a sterile condition. For this starch solution which containing ZnO NPs (100 mg concentration) was added into wells (6 mm width) on MHA plate. ZnO NPs embedded-starch solution was taken for the test because it is similar whether to test the film or to test the solution. The plate was incubated at 37$\pm$1$ ^\circ$C for 24 hrs. Antibacterial activity of ZnO NPs/starch solution was measured based on the zone of inhibition around the well infused with the rifampicin (RIF) and HPLC grade sterile water and biosynthesized silver nanoparticles/starch solution were used as positive and negative control respectively.
\end{enumerate}

\subsection{Qualitative Analysis of the ZnO Embedded Starch Film, Wrapped Over Grapes}
\begin{enumerate}
\item {\textbf{Color:}}
Variation in surface color of all vegetables was determined using calibrated Hunter Lab Colorimeter (Color Flex EZ, Hunter Lab, and USA) in terms of L*, a* and b* of each grapes in samples and then averaged all parameters.

\item {\textbf{Weight Loss:}}
Weight of all grapes in two batches (wrapped and non-wrapped) for 7days were measured. Weight loss (WL) of each vegetable in batches was calculated using the equation (\ref{eq3.8}): \\
\begin{equation}
WL(\%)=[(m_i-m_f/m_i]\times100  \label{eq3.8}
\end{equation}
Where, m$_i$ and m$_f$ is the initial an final weight of batches.

\item {\textbf{Total soluble solids (TSS):}}
Pulp was separated from grapes using sharp knife and then crushed in mortar and pestle. Thereafter, crushed pulped was squeezed using muslin cloth. Solid matter was discarded and juice was used for measurement of total soluble solids using Refractometer with range 0-32\%. Three replications/batches were performed and then findings were displayed in \% total soluble solids.

\item {\textbf{Titratable Acidity and pH:}}
Titratable acidity was measured using standard described method by A.O.A.C (1900). In brief, 6 g grapes juice was diluted in 25 mL distilled water in the conical flask. Diluted juice was titrated with 0.01 N sodium hydroxide solution using phenolphthalein (1\%) as an indicator. Light pink color of solution was observed at end of titration. Acidity was reported as g of citric acid/ 100 g of grapes pulp. pH was monitored during the storage period of grapes using a pH meter (Syntronics system-362, Ahmadabad, India).

\item {\textbf{Sensory Evaluation:}}
Sensory evaluation of coated and native samples of fruits was conducted using 9-point hedonic scale to check the quality and acceptability fruits after storage by a panel of 5 semi-trained members. Stored samples qualities and acceptability were judged on the basis of visual aspect, color, texture, flavor, and overall acceptability
\end{enumerate}

\section{Result and Discussion}
\subsection{Chemical Synthesis of ZnO Nanoparticles (NPs)}
\begin{enumerate}
\item {\textbf{X-Ray Diffraction (XRD) analysis:}}
XRD pattern of successfully synthesized ZnO NPs (after annealing) by sol-gel approach in the pH range of 8, 9, 10 and 11 are shown in Figure \ref{fig4.2}(a), which confirmed the wurtzite structure of the ZnO NPs {as peaks present in XRD pattern matches with JCPDS database card number [01-070-2551]}. The sizes of the ZnO nano-crystals at different pH ranges were calculated using Debye-Scherer formula which is shown in Table \ref{tab4.1}. The samples are highly crystalline with wurtzite phase and there is broadening in peaks due to high intensity extended over the 2$\theta$ scale correspond to lattice plane which are shown in Fig. \ref{fig4.2}(b). The broadening of the peaks gives an idea about the small particle size of the synthesized ZnO nanoparticles. The broadening in the ZnO peaks resulted due to lattice strain and lattice stress [as shown in Fig. \ref{fig4.2}(b)]. VD Mote \etal (2012), also explained the effect of strain and stress over the NPs \cite{r175}. Lattice strain also plays an important role in determining characteristics of nanoparticles, hence Williamson-Hall (W-H) Uniform Deformation Model (UDM) equation (\ref{eq4.1}) was used to evaluate average lattice strain as shown in Fig. \ref{fig4.2}(c).
\begin{equation}
\beta\cos\theta= K\lambda/D_{hkl} + 4 \epsilon\sin\theta  \label{eq4.1}
\end{equation}
It observed that crystal size calculated using Scherrer's and W-H equation are comparable with each other [as shown in Fig. \ref{fig4.2}(c)]. With increase in pH value from 8 to 10, crystallite size decreases rapidly due to increase in corresponding lattice strain and stress. Further increase in pH from 10 to 11, the crystallite size again increases but this is due to the increase in the alkanity of the solution. This behavior of size variation with pH is shown in Table \ref{tab4.1}. The average crystalline size of nanoparticles is calculated by Debye Scherrer's formula as shown in following Equation (\ref{eq4.2}):
\begin{equation}
D_{hkl} =  k\lambda/\beta\cos\theta  \label{eq4.2}
\end{equation}
where,\\
\begin{itemize}
\item D$_{hkl}$ is the crystalline size\\
\item $k$ is the structure constant or Scherrer's constant (which is 0.89 for spherical ZnO NPs)\\
\item $\lambda$ is the wavelength of the x-rays (Cu K$_{\alpha}$ radiation of 1.5418 \AA )\\
\item $\beta$ is the FWHM (Full Width Half Maxima)\\
\item $\theta$ is the diffraction angle or Bragg's angle\\
\end{itemize}

Cu K$_{\alpha}$ radiation of 1.5418 $\AA$ is accelerated on sample for single crystal diffraction while sample is rotating and intensity of reflected X-rays is getting recorded. Peak in intensity is occurred when constructive interference is occurred satisfying the Bragg Equation. X-ray signals are recorded and processed by the detector and is converted to count rate \cite{r180}, which makes the XRD pattern. The relative intensities and position of peaks obtained in diffraction profile is required for identification of phases from x-ray diffraction.

X-ray diffraction also helps in determining that how the presence of even millimeters sized defects or dislocations in crystals affects the diffraction peaks. For example, a small size of grain can change diffraction peak widths and very small crystals causes broadening of peak. The crystallite size is easily calculated as a function of peak width (specified as the full-width at half maximum peak intensity (FWHM)), peak position and wavelength.
\begin{figure}[htb!]\centering
  \includegraphics[width=0.48\textwidth]{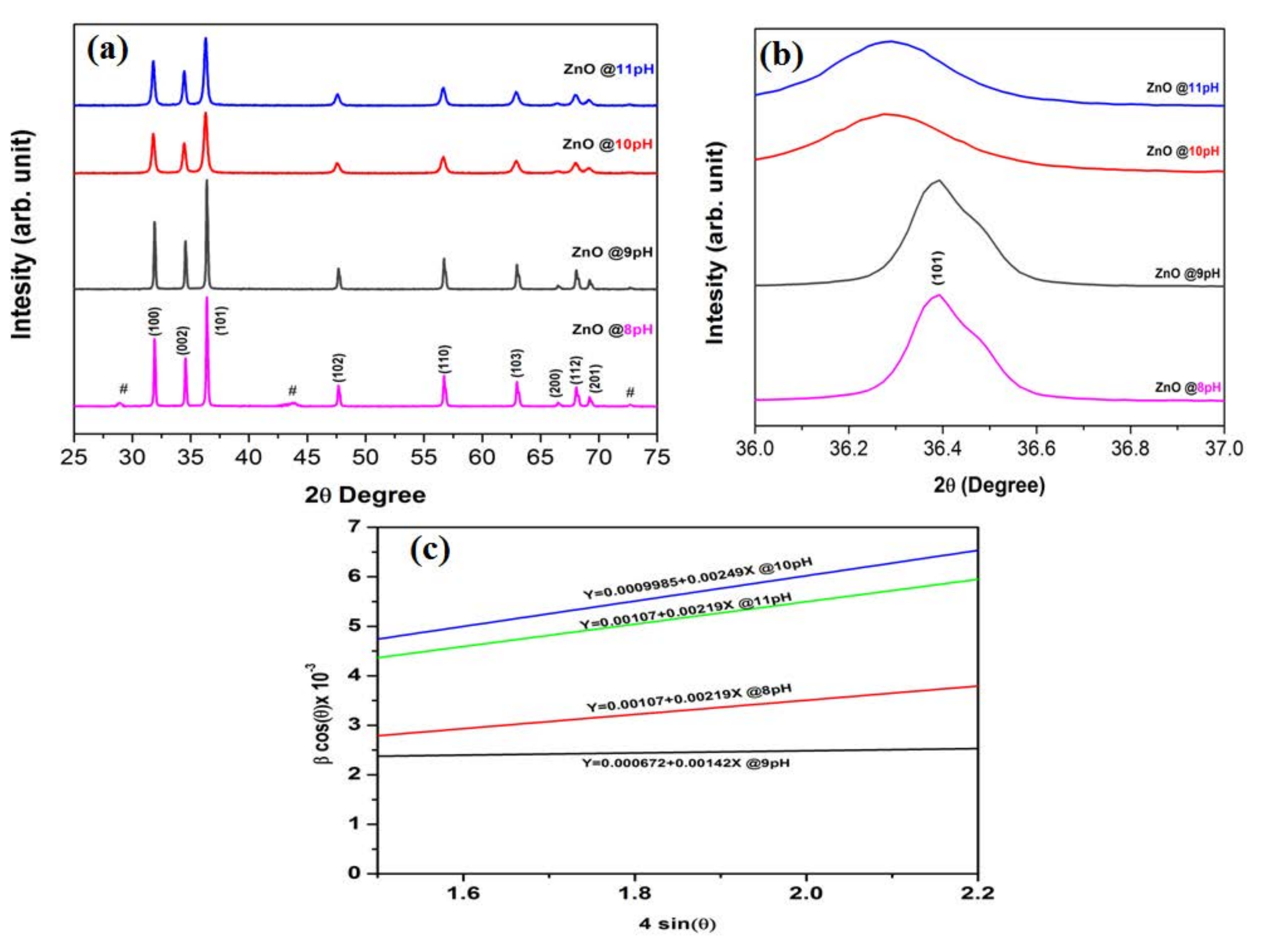}
  \caption{(a) XRD patterns at different pH range of 8 to 11 pH and impure phase in form of extra peaks marked with symbol \# for pH 8; (b) broadening and the lateral shift of prominent peaks at 2$\theta$= 36.399$^\circ$ at different pH level of 8, 9, 10 and 11 to the lattice plane of (101) and (c) showed W-H plots with linear fitting of lattice strain at different pH level of 8 to 11 pH.} \label{fig4.2}
\end{figure}
\begin{table}[htb!]\scriptsize{\centering
\caption{Crystallite size of ZnO NPs synthesized at different pH from 8 to 11.} \vskip 0.5cm
\begin{tabular}{|c|c|c|c|c|} \hline
pH value                   &  8pH     & 9pH      & 10pH  & 11pH     \\ \hline
Crystallite size           &          &          &       &          \\
(Debye Scherrer's formula) & 52.84 nm & 46.75 nm & 28 nm & 30.34 nm \\ \hline
\end{tabular} \label{tab4.1}}
\end{table}
With the change in pH value of the sol, hydrolysis and condensation also changes altering the structure, surface, size, and bond formation etc of ZnO nanoparticles. In addition, pH value controls H$^+$ and OH$^-$ in the solution thus determining the metal oxygen bond formation. The different hydrogen ion concentration is corresponds to the different size and shapes of nanoparticles formation. It has been investigated that ZnO nanoparticles generally forms in alkaline base since divalent metal ions don't hydrolyses in acidic nature. Increase in temperature and concentration of alkali increase the solubility of ZnO NPs in alkali medium, most commonly NaOH and KOH are used. Na$^+$ has smaller ion radius and thus leads to more chances of incorporating into ZnO lattice. Also, it forms capping around the ZnO crystal thereby restraining the growth of nanocrsytals, Xu \etal in 2011 \cite{r183}. Normally, the ZnO structure cannot be synthesized well at pH 6 because of the high concentration of H$^+$ ions and low concentration of OH$^{-}$ ions in the solution. For the neutral condition at pH 7, the number of H$^+$ ions reacting with number of OH$^{-}$ from NaOH is equivalent.

Further increasing the concentration of OH$^{-}$ from pH 9 reduces the ZnO crystallite and particle sizes because the amount of dissolved OH$^{-}$ was larger during synthesis of ZnO at pH $>$ 9. When ZnO reacts with too much OH$^{-}$, the dissolution of ZnO occurs \cite{r186}. This was due to preferential erosion of ZnO in acidic solution and the enhanced growth in neutral land alkaline solutions which favored the formation of Zn(OH)$_2$ which is an inter mediate molecule in the reaction to form ZnO \cite{r187}.

The peak position and intensities for nanoparticles system from pH 9 to 11 were well matched with the JCPDS database [01-070-2551] revealing the wurtzite phase formation. No other phase except wurtzite was observed in the XRD pattern. Further estimated values of lattice constants are characteristics of ZnO unit cell. On the other hand, at pH 8 diffraction peaks show a dissimilar behavior as along with ZnO, Zn(OH)$_2$ are also observed, marked with \#. However at pH 7, ZnO nanoparticles were not formed. With increasing pH, the rate of transformation of Zn(OH)$_2$  to ZnO occurs leading to the formation of intense pattern with single phase (as shown in Fig. \ref{fig4.2}(b)). Similar results were discussed by S.S. Aalias \etal (2010), in which ZnO NPs were synthesized at different pH level of 6 to 11. They observed at lover pH i.e. $<$ 7pH, ZnO NPs agglomerates easily but at higher pH i.e. $<$ 8 pH, fine powders were synthesized \cite{r60}.

\item {\textbf{Fourier Transform Infrared Ray (FT-IR) Spectroscopy Analysis:}}
Various functional groups attached to ZnO nanoparticles at different pH were determined by FTIR spectroscopy. In Fig. \ref{fig4.3}, the band at 3328.75-3336.10 cm$^{-1}$ is ascribed to O-H vibration group on the surface of ZnO nanoparticles. The peaks in region 612.38-625.09 cm$^{-1}$ and 429.53-431.84 cm$^{-1}$is due to ZnO stretching that showed the evolution of Zn(OH)$_2$ to ZnO.

\begin{figure}[htb!]\centering
  \includegraphics[width=0.48\textwidth]{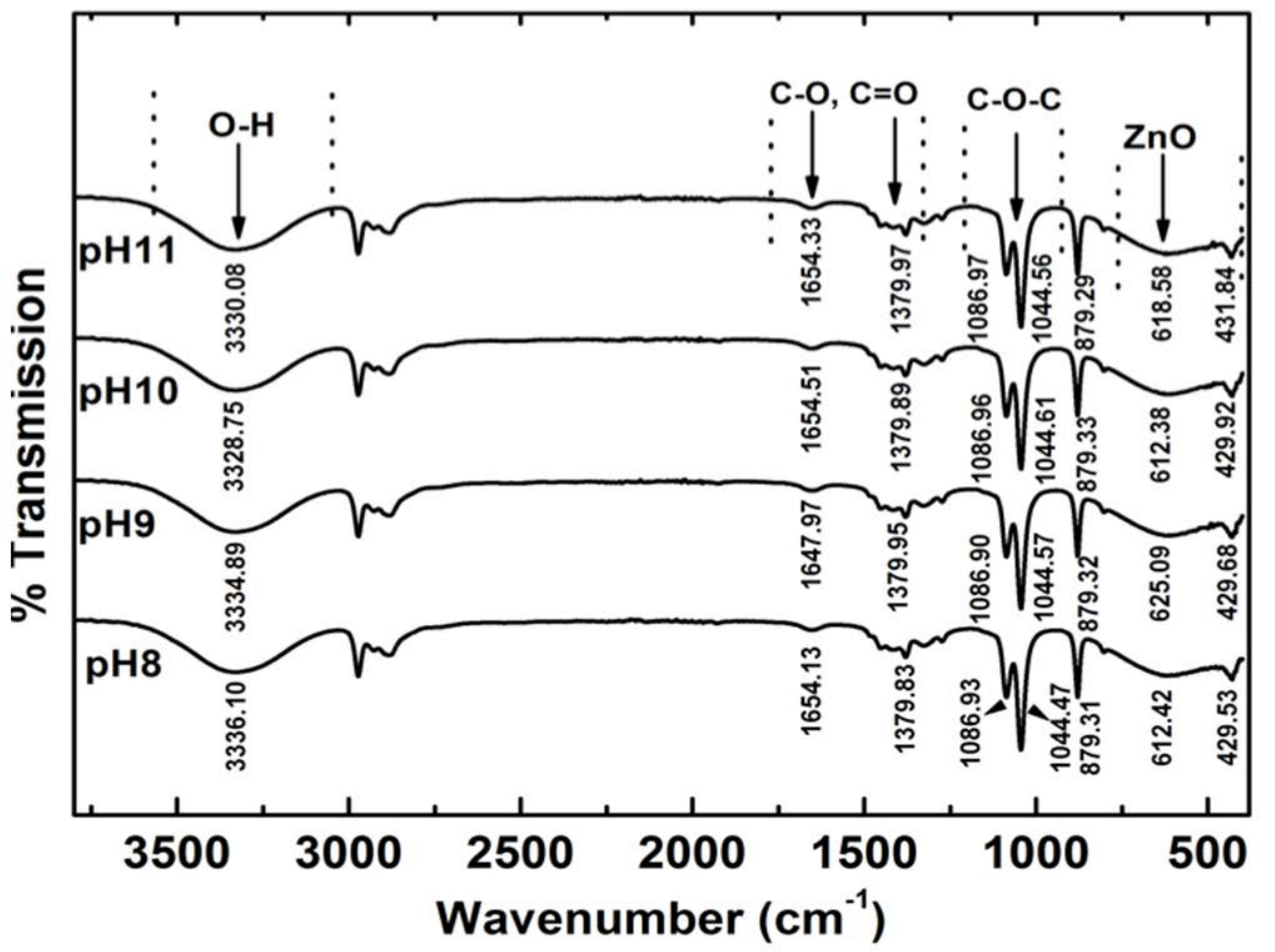}
  \caption{FTIR spectra of ZnO NPs at different pH range of 8 to 11, mentioning the band position and function and functional groups attached.}\label{fig4.3}
\end{figure}

Sharma \etal reported small ZnO peaks at 917 cm$^{-1}$and 676 cm$^{-1}$ \cite{r60}, whereas R. Wahab \etal marked ZnO peaks in the region 464-419 cm$^{-1}$. It is observed that the variation in wavenumber is due to different particle size on varying pH that caused shift in the IR peaks. The stretching in the region 1647.97-1654.33 cm$^{-1}$ , 1379.83-1379.97 cm$^{-1}$ and 1044.47-1086.90 cm$^{-1}$ is because of C-O, C=O and C-O-C functional groups (as shown in Fig. \ref{fig4.3} and Tab. \ref{tab4.2}). Presence of bands corresponding to these functional groups is common in nanoparticles synthesis using chemical method and is associated with organic matter left in the reaction. Similar results were shown by S.S. Aalias \etal (2010), where ZnO NPs peaks in FTIR results were at 450 to 550 cm$^{-1}$ \cite{r60}.

\begin{table}[htb!]\centering
\caption{Characteristic FTIR absorption peaks observed in Zinc Oxide nanoparticles synthesized at different pH varying from 8 to 11 using sol-gel route.} \vskip 0.5cm
\begin{tabular}{|c|l|l|} \hline
pH &	Wavenumber            & Functional group \\ \hline
8  & 3336.10, 1654.13, 1379.83,& O-H, C-O, C=O, \\
   & 1086.93, 429.53-612.42    &  C-O-C, ZnO resp.               \\ \hline
9  & 3334.89, 1647.97, 1379.95,&  O-H, C-O, C=O,  \\
   &1086.90, 429.68-625.09 & C-O-C, ZnO resp.\\ \hline
10 & 3328.75, 1654.51, 1379.89, & O-H, C-O, C=O, \\
   &1086.96, 429.92-612.38 &  C-O-C, ZnO resp.\\ \hline
11 & 3330.08, 1654.33, 1379.97, &  O-H, C-O, C=O,\\
   &1086.97, 431.84-618.58 &  C-O-C, ZnO resp.\\ \hline
\end{tabular} \label{tab4.2}
\end{table}
\item {\textbf{Ultra Violet-Diffusive Reflection Spectroscopy (UV-DRS) Analysis:}}
Fine powders were obtained when the pH of the sols was increased to 9 to 11. The maximum crystallite size of 55.84 nm was obtained at pH 8 and minimum size of 28 nm was obtained at 10 pH. The particles size of the ZnO synthesized between pH 8 and 11 were in the range of $~$28-55.84 nm as shown in the Tab. \ref{tab4.1}. Fig. \ref{fig4.7}(a \& b) in Ultraviolet-diffuse reflection spectroscopy analysis (UV-DRS analysis) demonstrated that ZnO NPs, synthesized from pH range from 7 to 11 pH has good optical properties with band gap energy (Eg) of 3.11 to 3.19 eV before annealed and 3.12 to 3.25 eV after annealed. S.S. Aalias \etal (2010) obtained similar results of band gap ranging from 3.14 to 3.25 eV at different pH range \cite{r60}. Variation in band gap of ZnO NPs (before and after annealing) with change in pH is shown in Fig. \ref{fig4.5}

\begin{figure}[htb!]\centering
  \includegraphics[width=0.22\textwidth]{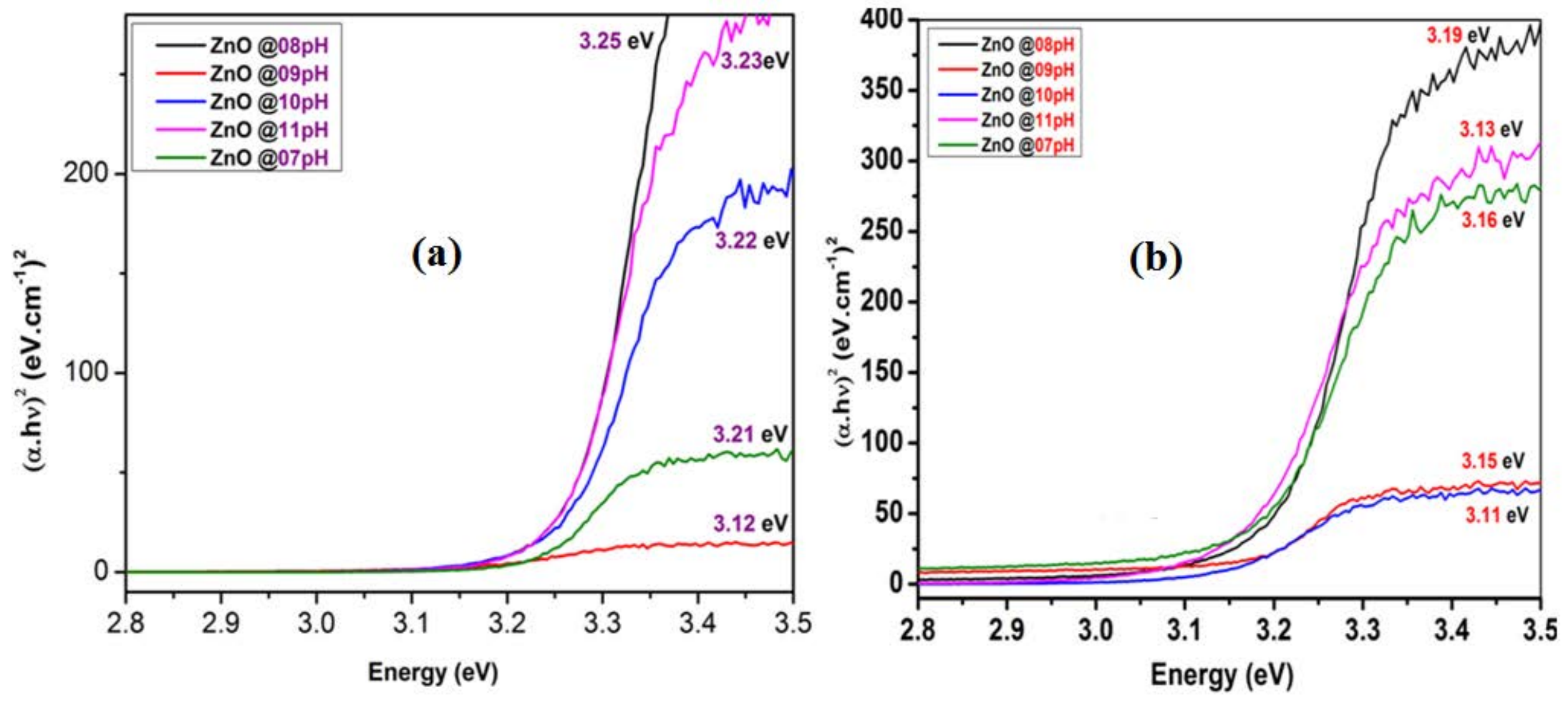}
  \caption{UV-DRS analysis (band gap) of ZnO NPs at different pH range of 7 to 11 pH (a) before \& (b) after annealed.}\label{fig4.7}
\end{figure}
\begin{figure}[htb!]\centering
  \includegraphics[width=0.48\textwidth]{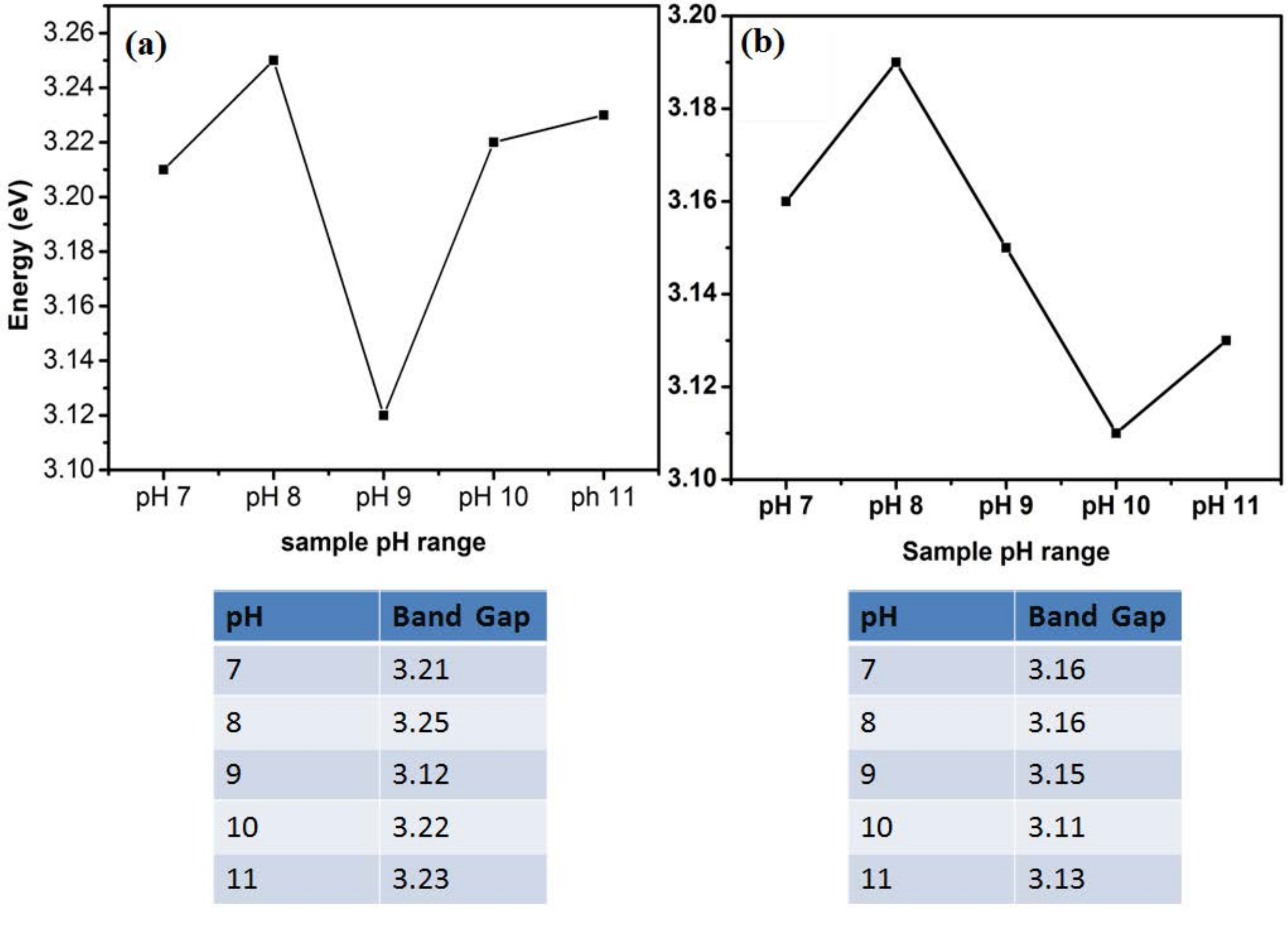}
  \caption{Effect of pH over the band gap of ZnO NPs; (a) before annealing \& (b) after annealing.}\label{fig4.5}
\end{figure}

\item {\textbf{Field Emission Scanning Electron Microscopy (FE-SEM) Analysis:}}
Based on the FE-SEM results (as shown in Fig. \ref{fig4.8}), it was observed that formation of hexagonal shape ZnO nanostructure could be achieved without use of any surfactant or catalytic agent. The growth process of ZnO nanostructures was supposed to be occurred in different phases i.e., nucleation of nanoparticles and crystal growth.
\begin{figure}[htb!]\centering
  \includegraphics[width=0.48\textwidth]{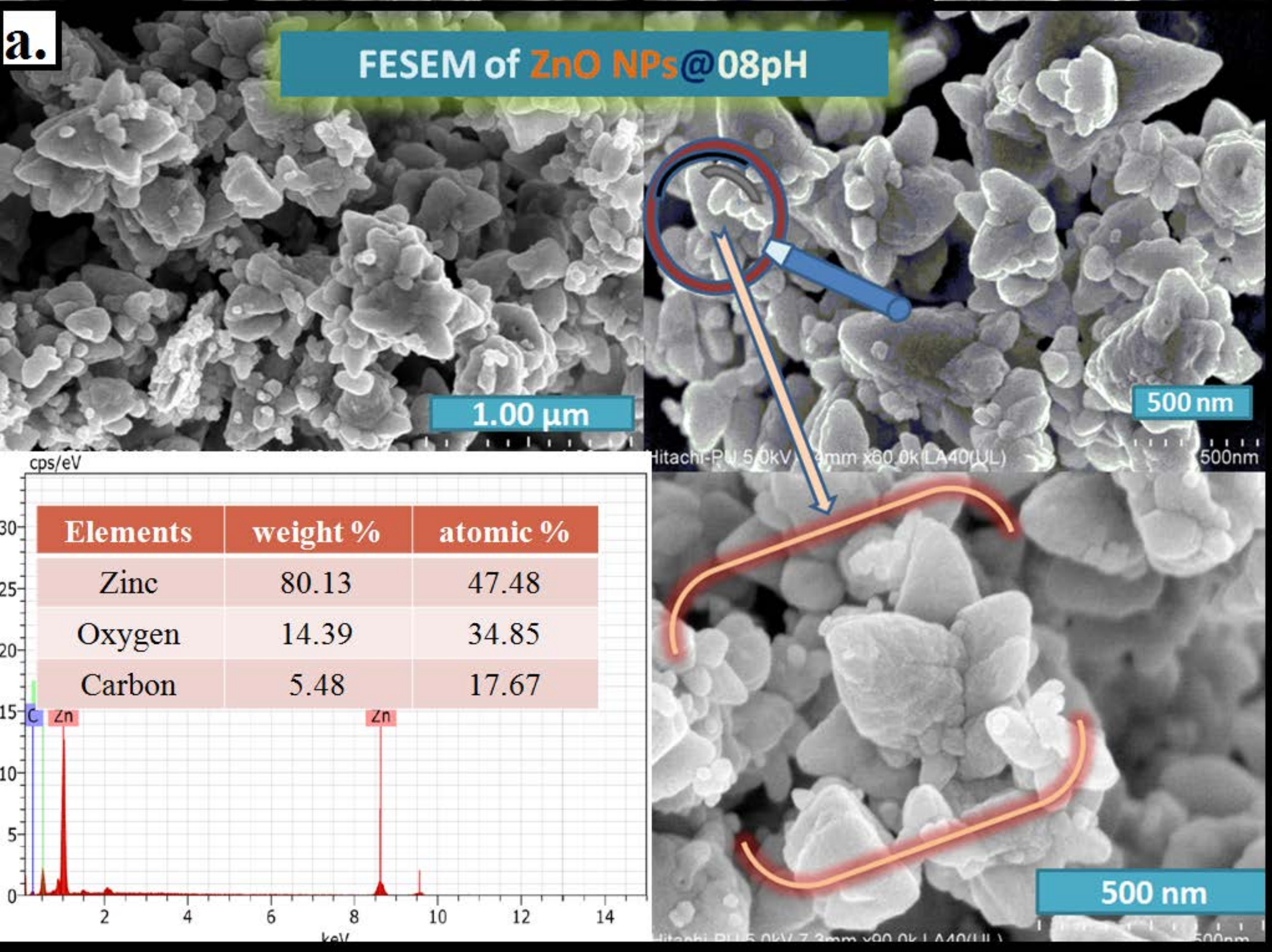} \vskip 0.5cm
  \includegraphics[width=0.48\textwidth]{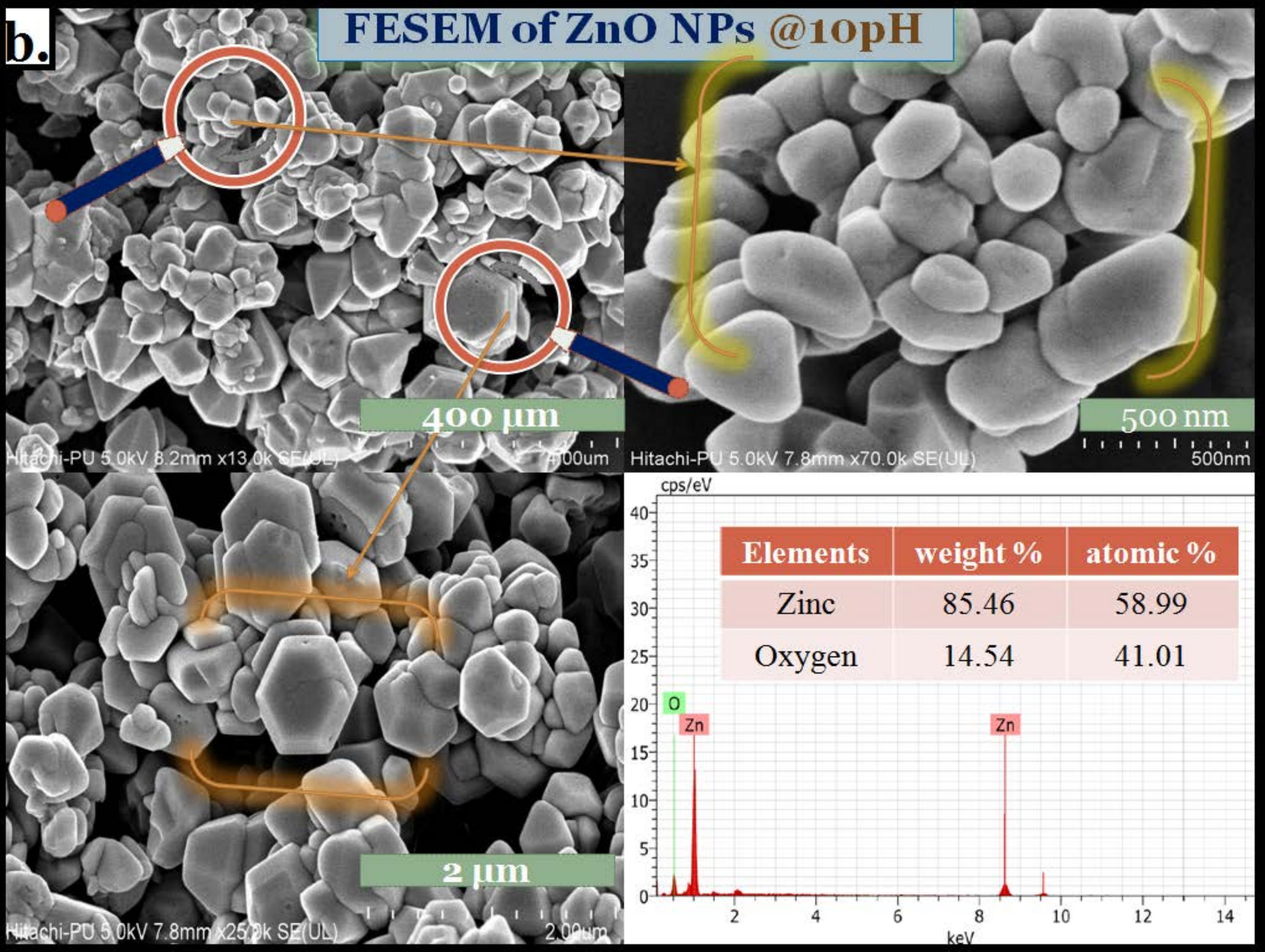}
  \caption{FE-SEM images of ZnO NPs after annealed at (a) 8pH and (b) 10pH.}\label{fig4.8}
\end{figure}
The crystallite size was calculated using Debye Scherrer's formula as explained in equation (\ref{eq4.2}) in XRD analysis. It has been observed that with increase in pH value from 8 to 11 pH, the crystallite size of the nanoparticles also gets changed (as shown in Tab. \ref{tab4.1}. The particle size and their morphology were studied by FE-SEM of ZnO nanoparticles as shown in Figs. \ref{fig4.8}(a) and \ref{fig4.8}(b). In Fig. \ref{fig4.8}(a) the ZnO NPs synthesized at 8 pH showed the non-uniformity in the structural morphology in ZnO NPs with some extent of impurities in the form of extra peaks which also detected in XRD pattern [as shown in Figs. \ref{fig4.2}(a) and \ref{fig4.2}(b),with symbol \#].This may be due to the number of H$^+$ ions reacting with number of OH$^{-}$ from KOH or NaOH was nearly equivalent. It has been investigated that ZnO nanoparticles generally forms in alkaline base since divalent metal ions doesn't hydrolyses in acidic nature \cite{r15}. ZnO NPs synthesized at 10pH indicated hexagonal and spherical structures with overall uniformity in the particle size.

FE-SEM of 8 pH and 10 pH was done to show the difference between the crystallinity structures. Energy-Dispersive X-ray Spectroscopy (EDX) was done to evaluate the overall chemical composition of synthesized nanoparticles. EDX of synthesized ZnO NPs at 8pH and 10pH was done because of their maximum and minimum nano-crystal sizes (as shown in Tab. \ref{tab4.1}. It was clearly observed that at 8pH mixture of three elements [as shown in Fig. \ref{fig4.8}(a)] i.e. Zinc (Zn), Oxygen (O) and Carbon (C) with atomic \% of 47.48, 34.85 and 17.67 respectively were seen through EDX but in case of 10 pH we observed only mixture of two elements with higher atomic \% [as shown in Fig. \ref{fig4.8}(b)]. We observed weight \% and atomic \% ratios with EDX.  Atomic ratio of Zinc to Oxygen in ZnO NPs is 1.36:1 at 8 pH but at 10 pH atomic ratio is 1.5:1 which is comparatively higher than the first one. During the initial stage of the reaction, the aqueous solution of Zn(CH$_3$COO)$_2$.2H$_2$O reacts with KOH to form zinc hydroxide [Zn(OH)$_2$], potassium acetate (CH$_3$COOK) and water molecules.
Following equation (\ref{eq4.3}) describes the growth mechanism of ZnO:
\begin{equation}
Zn(Ac).2H_2O+2KOH\Longleftrightarrow Zn(OH)_2+2K(Ac)+2H_2O  \label{eq4.3}
\end{equation}
The build-up of precursor molecules results in a super-saturation of the solution. Nucleation occurs continuously above and below the condensation and polymerization threshold. The Zn(OH)$_2$ reacts with the water molecule to form the growth unit Zn(OH)$_2$ and hydrogen ions (2H$^{2+}$) in as shown in equation (\ref{eq4.4}):
\begin{equation}
Zn(OH)_2+ 2H_2O\Longleftrightarrow Zn(OH)_4^{2-}+ 2H^{2+}  \label{eq4.4}
\end{equation}

The further growth of ZnO and separation of agglomerates from the supersaturated solution is possible until the solution becomes saturated with a white precipitate of colloidal gel Zn(OH)$_4^{2-}$. Centrifugation transforms the Zn(OH)$_4^{2-}$ into ZnO according to the following equation (\ref{eq4.5}):
\begin{equation}
Zn(OH)_4^{2-}\Longleftrightarrow ZnO + H_2O + 2OH^-  \label{eq4.5}                                                                                 \end{equation}

Higher OH$^-$ concentrations i.e. at pH 10 and 11, ZnO reacts with more OH$^-$ to form more amount of Zn(OH)$_4^{2-}$. These growth units easily form linkage on the ZnO surface (undergo agglomeration) and hence each crystal grows to square/pyramid shape structure. The dissolution of ZnO during reverse reaction produce Zn(OH)$_4^{2-}$ according to equation(\ref{eq4.6}):
\begin{equation}
ZnO + H_2O + 2OH^-\Longleftrightarrow Zn(OH)_4^{2-} \label{eq4.6}                                                                                     \end{equation}

To conclude, nano ZnO synthesized by Sol-Gel method it observed that the crystallite size of the ZnO NPs was highly affected by pH of the working solution. Increasing pH from 8 to 10 decreases the crystallite size whereas further increase in pH from 10 to 11 leads to increase in the crystallite size (which led to increase in particle size of the ZnO NPs). The variation of band gap with crystallite size is given in Fig. \ref{fig4.7}(a) and \ref{fig4.7}(b), which showed that ZnO with good optical properties which also affected by various pH range. The pH variation effect the synthesis of the ZnO NPs. From the precipitation profile, it was observed that increasing pH led to shortened precipitation time, which also means increasing particle size as seen in higher pH of 11 as shown in Fig. \ref{fig4.2}. Similar results were discussed by S.S. Aalias \etal (2010) \cite{r60} and S. Xu \etal (2011) \cite{r183}.
\end{enumerate}
\subsection{Green synthesis of ZnO Nanoparticles}
Through visual observation the qualitative analysis was carried out. The white precipitation formation as the end product of the solution, gives the confirmation of the synthesis of the ZnO NPs. Some reducing sugars such as flavanones, terpenoids etc. which are the main constituent of the neem extract acts as the stabilizing agents. Aldehyde group which present inside the neem broth plays an important role in reduction of the Zinc ion to their respective zinc oxide NPs. These phytochemicals also acts as stabilizing agents \cite{r140,r141}.
\begin{enumerate}
\item {\textbf{Ultraviolet-Visible (UV-VIS) Spectroscopy Analysis:}}
It is done to check whether the ZnO NPs synthesized or not. Neem extract acts as a natural stabilizing and reducing agent which reduces the zinc acetate dihydrate salt putted inside the extract to ZnO (as shown in Fig. \ref{fig4.9}). Potassium hydroxide (KOH) was used drop wise for raising the pH ($~$11 pH). After getting white precipitate the magnetic stirrer stayed on for 2-3 hrs and UV-vis is done to cross check the confirmation of the formation of the ZnO NPs as shown in Fig. \ref{fig4.10}. The absorption peak appeared at $~4$380 nm in 2\% neem extract and $~$380.12 nm in 3\% neem extract confirms the synthesis of ZnO NPs. The comparison of both 2\% and 3\% neem extract containing synthesized ZnO NPs through green approach was shown in Fig. \ref{fig4.9}(c). Jinxia Ma \etal (2016) \cite{r121}, explained that the pure ZnO NPs has absorption peak at 380 nm. The shift of the peak was possibly due to the presence of the neem extract in high concentration. Centrifuge is done at 5000 rpm for 3 minutes and washed several time (nearly 3 to 4 times) with absolute ethanol to remove the water soluble impurities present inside the sol.
\begin{figure}[htb!]\centering
  \includegraphics[width=0.48\textwidth]{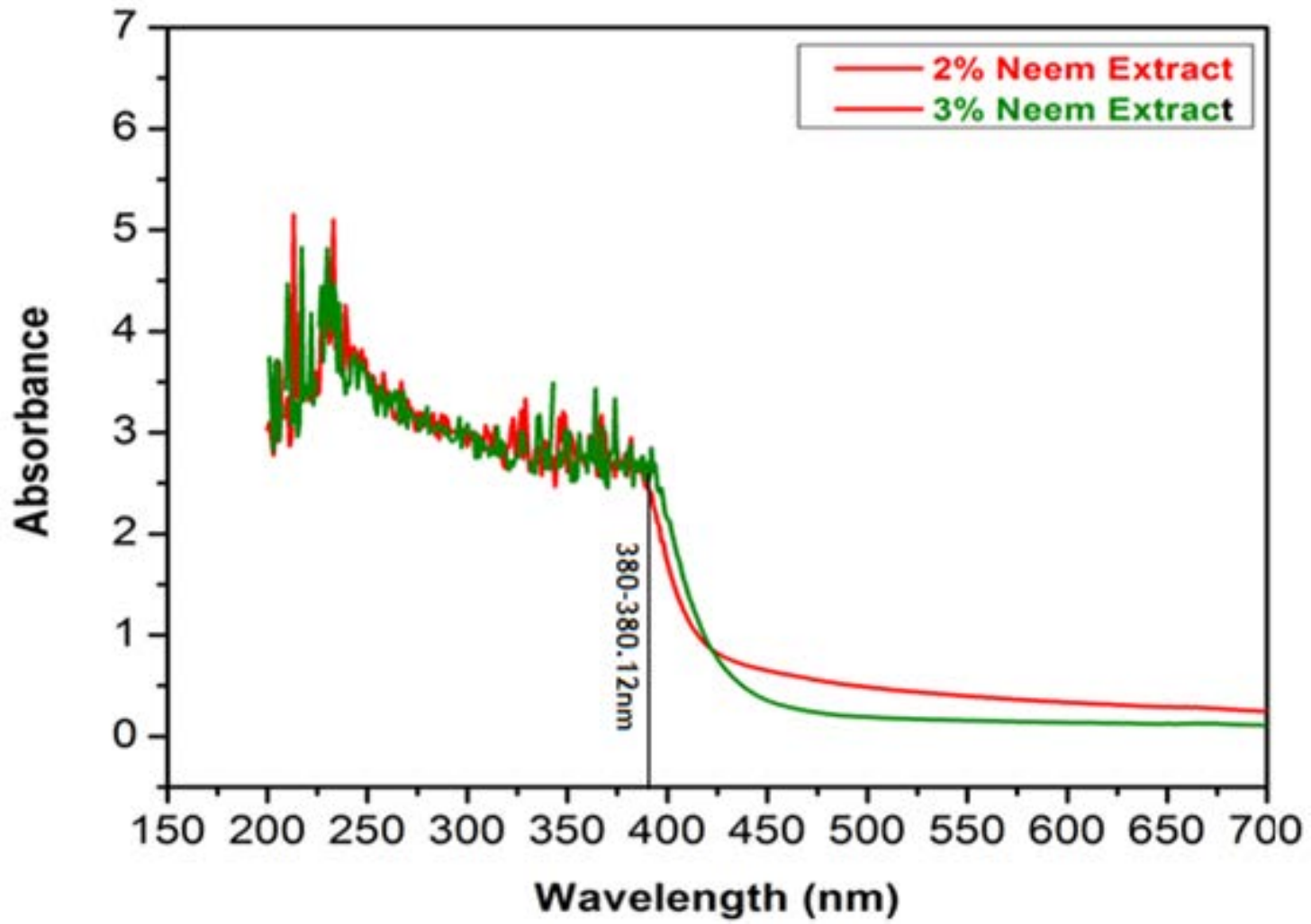}
  \caption{UV-vis of of solution containing 2\% and 3\% neem extract which also containing ZnO nanoparticle precursor solution.}\label{fig4.9}
\end{figure}

\item {\textbf{X-Ray Diffraction (XRD) analysis:}}
XRD pattern of successfully synthesized ZnO NPs by using green extract of \emph{A. indica} (Neem) is shown in Fig. \ref{fig4.10}(a). Sharp peaks in XRD pattern shows high crystallinity of ZnO without any presence of impurities in the form of extra peaks; as all the peaks present in XRD pattern matches with JCPDS card no. [01-070-2551]. The XRD pattern of ZnO NPs synthesized by neem extract at the different concentration of 2\% and 3\% are indicated by B1 and B2 shown in Fig. \ref{fig4.10}(a). It was observed that the concentration with 3\% and 2\% neem extract gives the crystallite size of 32.68 and 36.34 nm respectively which was calculated by Debye Scherrer's formula (as shown in Tab. \ref{tab4.3}). The broadening in the peak is zoomed out in Fig. \ref{fig4.10}(b) and also shown that ZnO NPs synthesized by 3\% neem extract has more broadening in XRD peaks as compare to the ZnO NPs synthesized by 2\% neem extract.
\begin{figure}[htb!]\centering
  \includegraphics[width=0.16\textwidth]{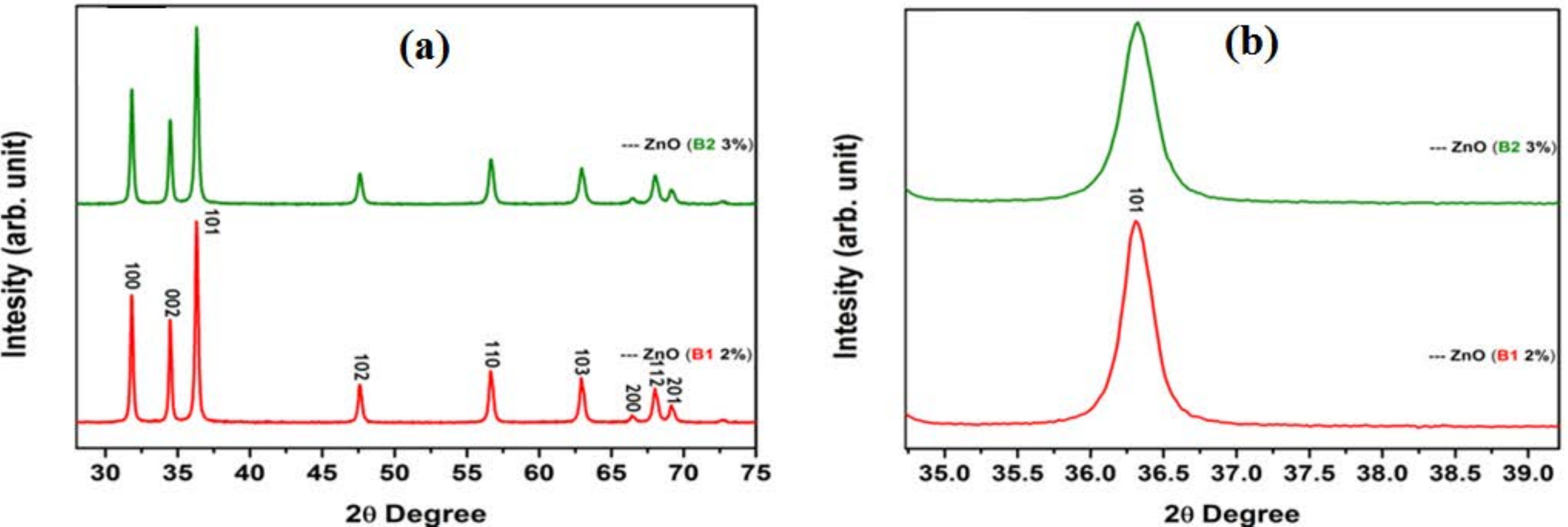}
  \caption{XRD pattern of pure ZnO NPs synthesized by using green extract of neem at different concentration of 2\% and 3\% which is indicated by B1 and B2 (a) indicated the comparison of green synthesized ZnO NPs XRD peaks and (b) broadening of the prominent peaks at 2$\theta$ of 36.308$^\circ$ to the lattice plane of (101).} \label{fig4.10}
\end{figure}
The broadening in the prominent peak [which shown in Fig. \ref{fig4.10}(b)] was due to the lattice strain (as shown in Fig. \ref{fig4.11} by W-H plot) and lattice stress. W-H plot also plays an important role in determining the characteristics of nanoparticles; hence Williamson-Hall (W-H) was used to evaluate the particle size by the lattice strain inside the particle [as shown in Fig. \ref{fig4.11}].
\begin{figure}[htb!]\centering
  \includegraphics[width=0.27\textwidth]{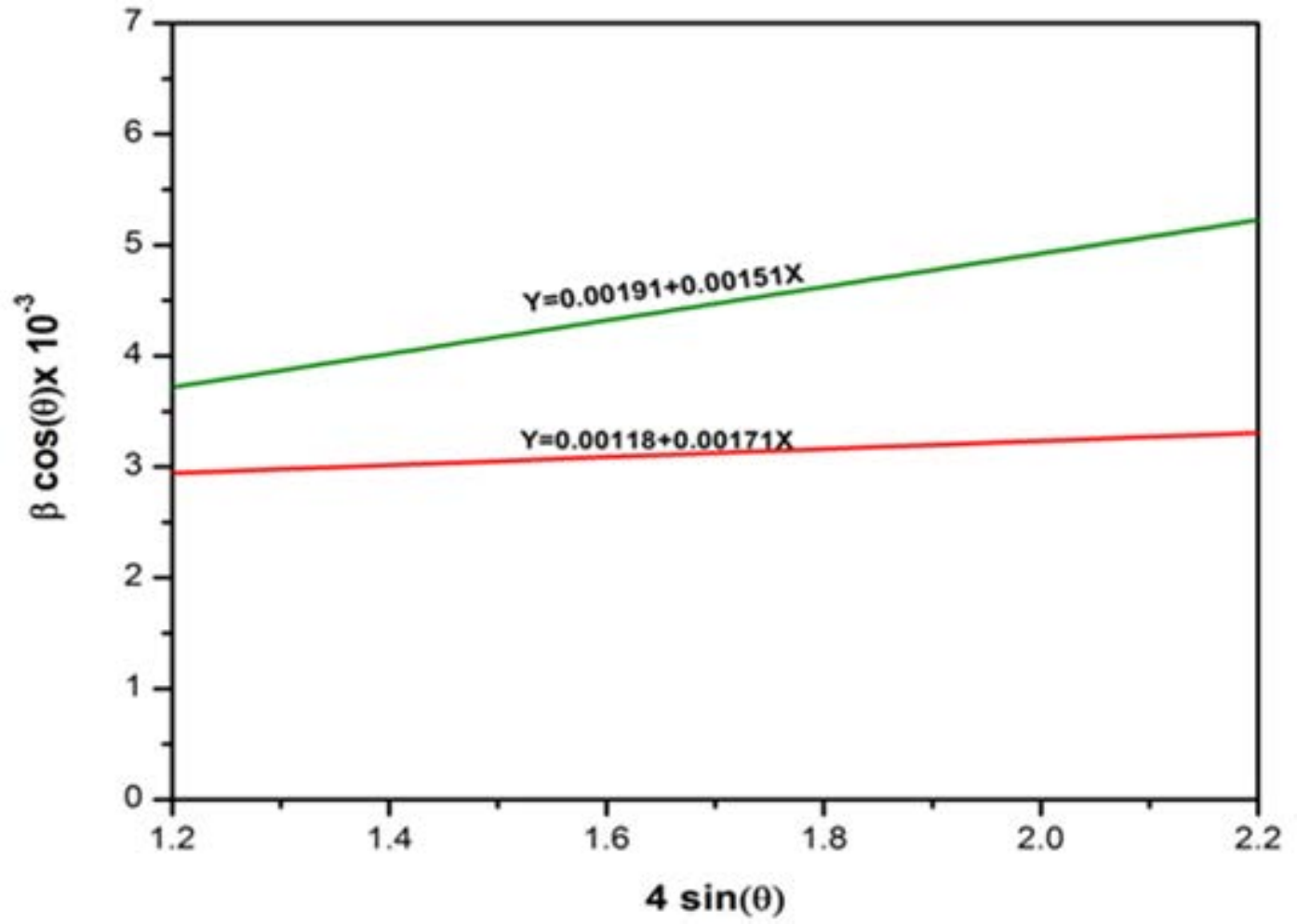}
  \caption{W-H plot for synthesized ZnO NPs at 2\% neem extract and 3\% neem extract.}\label{fig4.11}
\end{figure}

It was clear that with increase in lattice strain the broadening in the prominent peak also get increases; which ultimately results into the decrease in the size of the ZnO nanoparticles as per Figs. \ref{fig4.10}(b), \ref{fig4.11} and Tab. \ref{tab4.3}. Tmamanna Bhuyan \etal (2015), synthesized the ZnO NPs with same procedure \cite{r140}.
\begin{table}[htb!]\centering
\caption{Crystallite size of ZnO NPs synthesized at different pH from 8 to 11.} \vskip 0.5cm
\begin{tabular}{|c|c|c|} \hline
Neem extract               & B1 (2\%) & B2 (3\%) \\ \hline
Crystallite size           &          &          \\
(Debye Scherrer's formula) & 36.34 nm & 32.68 nm \\ \hline
\end{tabular} \label{tab4.3}
\end{table}

\item {\textbf{Fourier Transform Infrared Ray (FT-IR) Spectroscopy Analysis:}}
Fig. \ref{fig4.12} and Tab. \ref{tab4.4}, illustrates the FT-IR spectra of ZnO NPs synthesized by using neem extract. It represents the major functional groups that present inside the ZnO NPs-Neem Extract spectra, which possibly plays an important role in reduction and stabilization of the NPs. The spectra showed the absorption peaks at 3342.8279-3450.2131 cm$^{-1}$, 2927.505-2920.9406 cm$^{-1}$, 1629.3517-163591636 cm$^{-1}$, 1382.358-1388.923 cm$^{-1}$, 1022.125-1028.689 cm$^{-1}$ (for B1 \& B2) indicating the presence of capping and stabilizing agents. B1 is for the plant extract containing neem extract of 2\% and B2 is for the plant extract containing neem extract of 3\%.
\begin{figure}[htb!]\centering
  \includegraphics[width=0.48\textwidth]{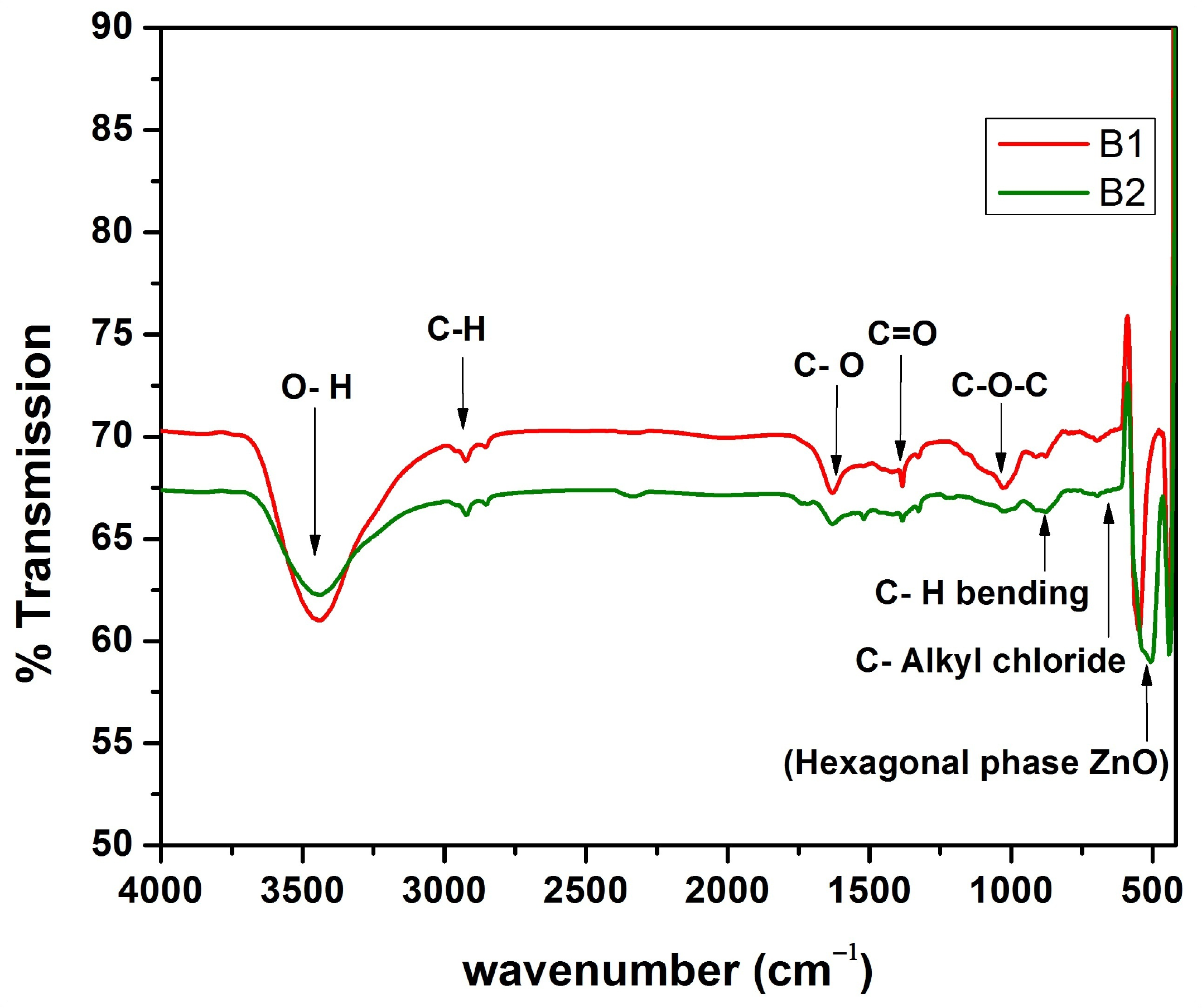}
  \caption{FT-IR spectrum of ZnO NPs synthesis by neem extract.}\label{fig4.12}
\end{figure}
The absorption broad peak of 3442.8279 and 3450.2131 cm$^{-1}$ indicated the presence of stretching vibrations of the O-H group which decreases with increase in the concentration of the neem extract. Peak range in 2927.505-2920.9406 cm$^{-1}$ corresponded to the group of C-H stretching \cite{r37}. Peak range 1629.3517-1635.9164 cm$^{-1}$ corresponded to the group of C-O stretching. The stretching in region 1382.358-1388.923 cm$^{-1}$, 1022.125-1028.689 cm$^{-1}$ is because of C=O and C-O-C functional groups. 650.0100-652.7654 cm$^{-1}$ corresponded to the C-Alkyl chloride and 528.1368-556.03643 cm$^{-1}$ corresponded to the hexagonal phase of ZnO nanoparticles \cite{r34}.
S. Azizi \etal in 2012; Tamanna Bhuyan \etal in 2015 and J. Santhoshkumar \etal in 2017; shown the similar results of FT-IR \cite{r137,r140,r195} as they all synthesized the ZnO nanoparticles with green synthesis approach by using plant extracts. Thus, through the FT-IR it is confirm that there is presence of many functional groups along with the ZnO nanoparticles.

\begin{table*}[htb!]\centering
\caption{FT-IR spectral peaks of ZnO NPs synthesized by neem extract.} \vskip 0.5cm
\begin{tabular}{|c|c|c|c|} \hline
S/No.& Absorption peak (cm$^{-1}$)& Absorption peak (cm$^{-1}$)& Functional  \\
     & in ZnO NPs(for B1)         & in ZnO NPs (for B2)        &             \\ \hline
1.	 & 3442.8279	              & 3450.2131	               & O-H         \\ \hline
2.	 & 2927.5050	              & 2920.9406	               & C-H         \\ \hline
3.	 & 1629.3518	              & 1635.9164	               & C-O         \\ \hline
4.	 & 1382.3580	              & 1388.9227	               & C=O         \\ \hline
5.	 & 1022.1250	              & 1028.6891	               & C-O-C       \\ \hline
6.   & 650.0100                   & 652.7654                   & C-Alkyl chloride \\ \hline
7.   & 556.0364                   & 528.1368                   & Hexagonal phase ZnO \\ \hline
\end{tabular} \label{tab4.4}
\end{table*}
\item {\textbf{Field Emission-Scanning Electron Microscopy (FE-SEM) Analysis:}}
It is observed that the concentration with 3\% (i.e. B2) and 2\% (i.e. B1) neem extract gives the crystallite size of 32.68 nm and 36.34 nm respectively which is calculated by Debye Scherrer's formula (as shown in Tab. \ref{tab4.3}). FE-SEM of only B2 was done because of its XRD results, which gives the crystallite size of ZnO nanoparticles which was nearly 32.68 nm. Through FE-SEM (as shown in Fig. \ref{fig4.13}) it was observed that most of the particles are of spherical and some are of hexagonal in shapes.
\begin{figure}[htb!]\centering
  \includegraphics[width=0.48\textwidth]{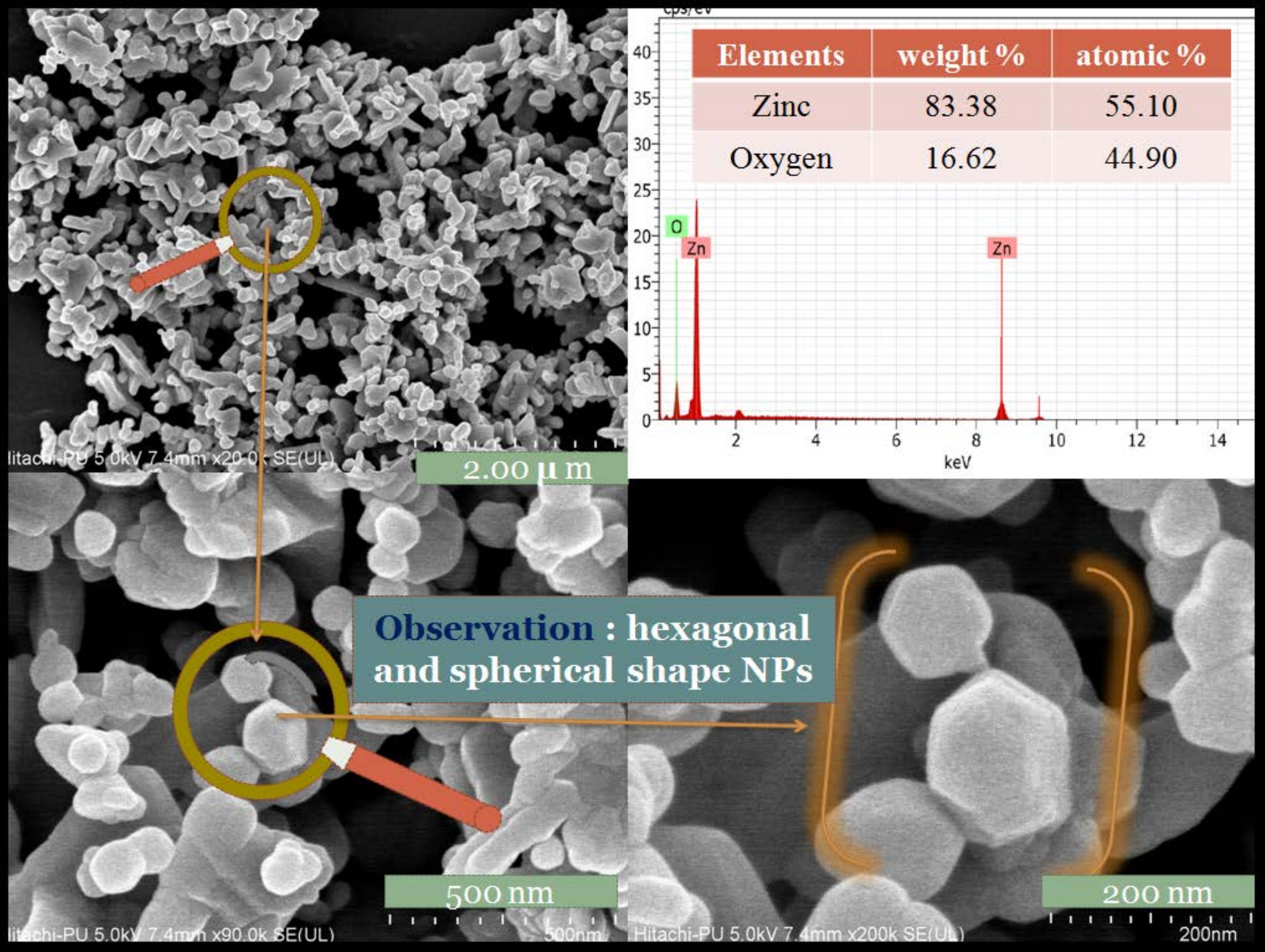}
  \caption{FE-SEM image of ZnO NPs synthesized by green approach at 10pH.}\label{fig4.13}
\end{figure}
There is overall homogeneity of particles with spherical and hexagonal with an average particle size of nearly 32.68 nm. Agglomeration of some particles is also seen at 2$\mu$m resolution range and it may be due to the annealing effect of ZnO nanoparticles at high temperature. Zoomed view of some location is also shown in figure just to show the hexagonal structure which is also developed inside the particles distribution of the ZnO nanoparticles. Similar result were shown by T. Bhuyan \etal in 2015 \cite{r140} and Elumalai \etal in 2015 \cite{r145,r196} who synthesized the same ZnO nanoparticles through green synthesis approach. Energy-Dispersive X-ray Spectroscopy (EDX) was done to evaluate the overall chemical composition of synthesized nanoparticles. Here in Fig. \ref{fig4.13} the presence of Zinc(Zn) and Oxygen(O) composition was calculated through the intensity of the peaks in EDX spectra, which equals the composition of same elements in ZnO.  It shows the weight\% and the atomic\% of the two elements. The atomic ratio of Zn to O is 1.22:1 which is very good indication for the ZnO composition.
\end{enumerate}

\subsection{Development of ZnO Embedded Corn Starch Film}
\begin{figure}[htb!]\centering
  \includegraphics[width=0.48\textwidth]{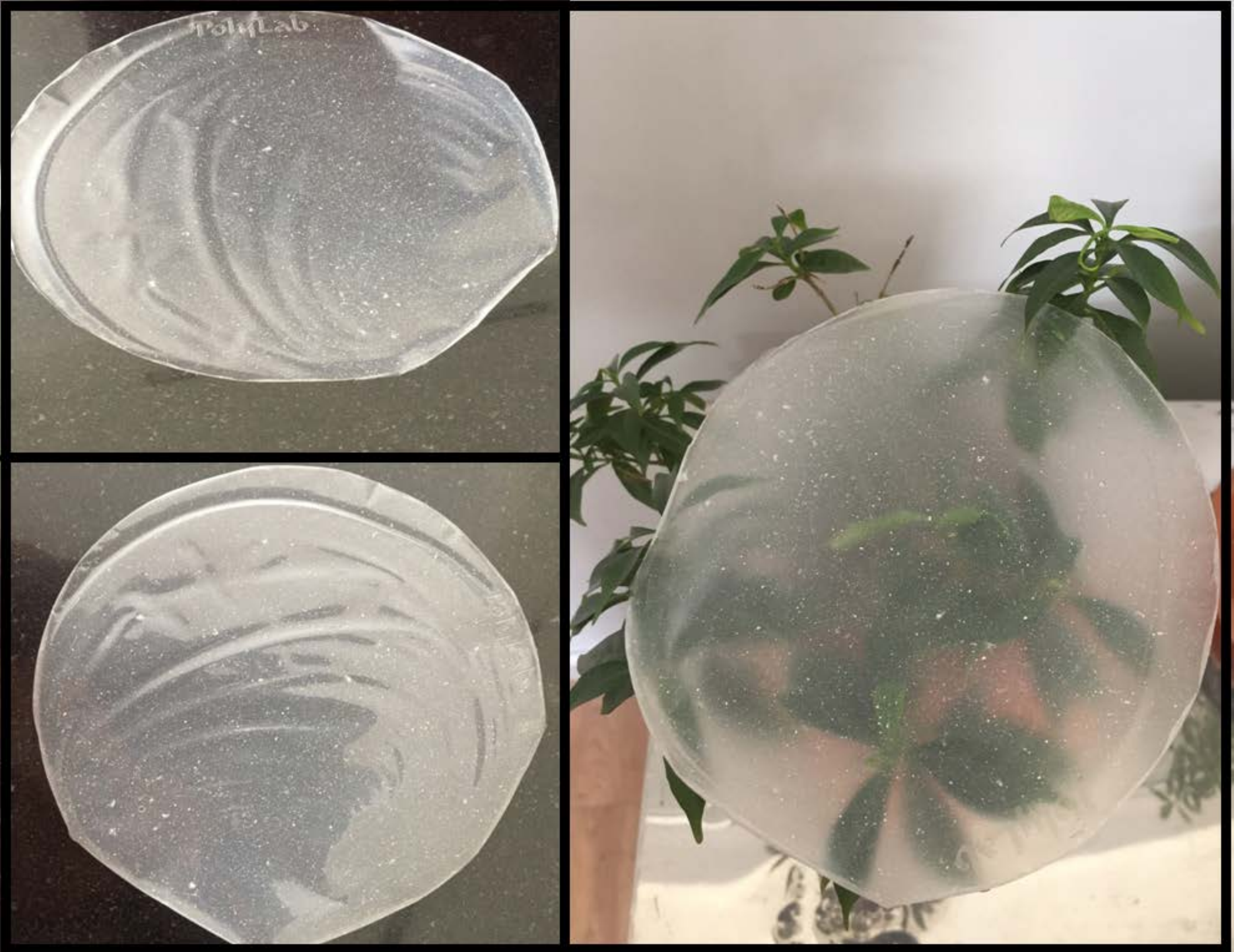}
  \caption{Chemically synthesized ZnO NPs (at 10pH) embedded corn starch film (F1).}\label{fig4.14}
\end{figure}

\begin{figure}[htb!]\centering
  \includegraphics[width=0.48\textwidth]{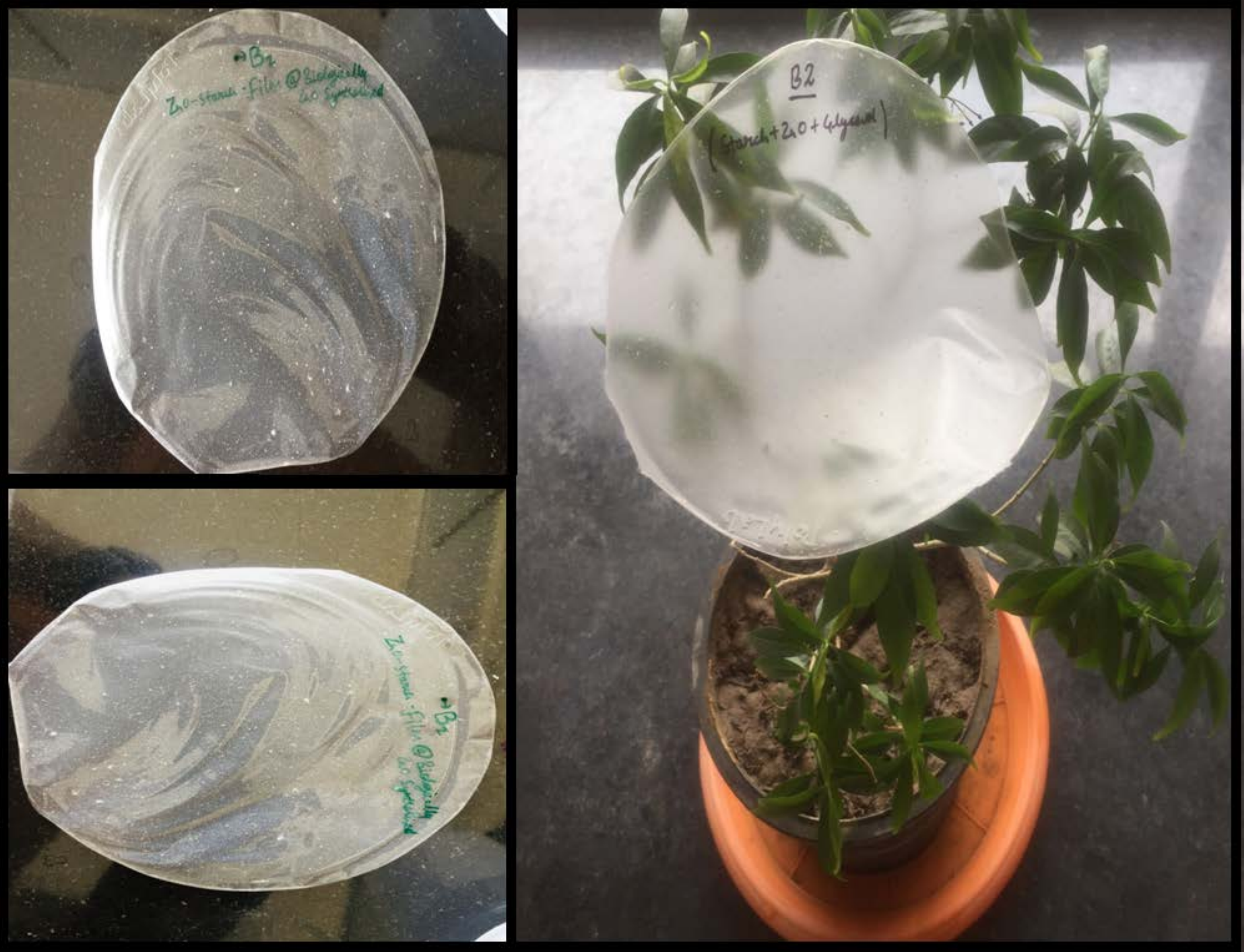}
  \caption{Biologically synthesized ZnO NPs (at pH10) embedded corn starch film (F2).}\label{fig4.15}
\end{figure}

\begin{figure}[htb!]\centering
  \includegraphics[width=0.48\textwidth]{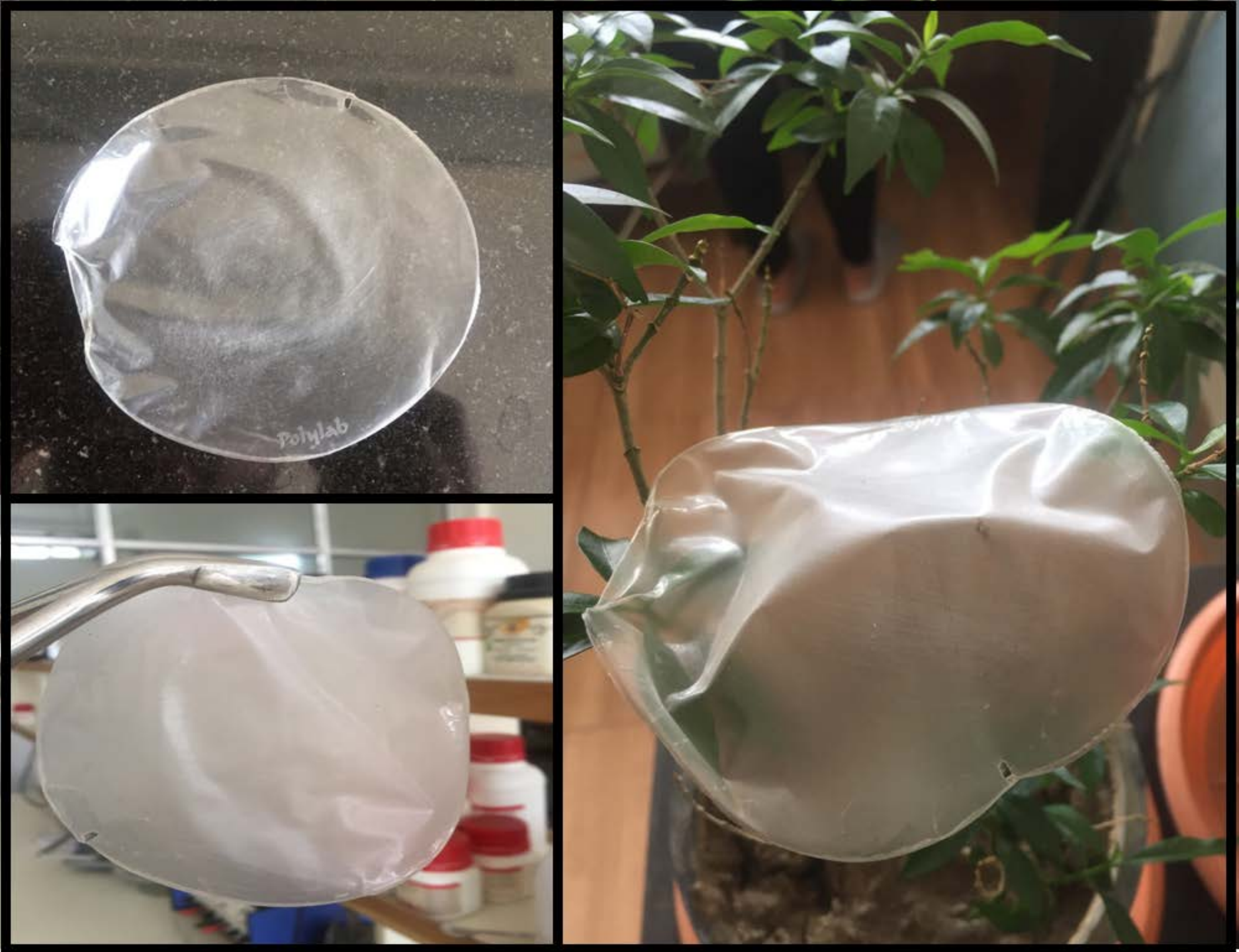}
  \caption{Corn-starch film (F3); containing (corn starch, acetic acid as cross-linker and glycerol as plasticizer).}\label{fig4.16}
\end{figure}

In order to enhance the properties of the nanocomposites, the choice of the polymeric materials as well as the homogeneous dispersion of the well sized and shaped NPs is needed. Enhanced properties of packaging materials are generally achieved when small distribution of NPs is attained \cite{r186,r187,r165}. However, the main limitations of this approach are the development of ZnNPs clumps and the heterogeneous distribution of NPs within the polymeric network. In this study, green method and chemical method was employed to prepare ZnO NPs which further utilized in the synthesis of the nanocomposites films (as shown in Fig. \ref{fig4.14} and Fig. \ref{fig4.15}). Corn starch film without containing any nanoparticles it is, is shown in Fig. \ref{fig4.16}.

\subsection{Analysis of Functional Properties of the Synthesized Films}
The functional properties and characterization of corn starch-based nanocomposites films or ZnO NPs embedded films were categorized into three parts as shown in Tab. \ref{tab4.5}. All testing of the film is done after putting it under the controlled environment condition of temperature and humidity ($~$75\%) inside a dedicator.
\begin{table}[htb!]\centering
\caption{Different types of synthesized films.} \vskip 0.5cm
\begin{tabular}{|c|l|} \hline
Films & Film characteristics \\ \hline
F1    &Chemically synthesized ZnO NPs at pH 10 embedded\\
      &corn starch film in which smallest calculated crystal \\
      & size of 28 nm of ZnO \\
      &NPs is taken (Table 4.1) \\ \hline
F2	  &Biologically synthesized ZnO NPs embedded corn \\
      &starch film (with B2: 32.68 nm crystallite size ZnO \\
      &NPs at maintained pH 10) \\ \hline
F3	  & starch film without embedded any NPs\\ \hline
\end{tabular} \label{tab4.5}
\end{table}

\begin{enumerate}
\item {\textbf{Film thickness(FT):}}
Thickness of the films was measured using Micrometer (Mitutoyo, Japan). The measurements were taken at 10 random positions of the films and average value was considered. The thickness of the film directly links with microstructure and degree of orientation of molecules in the films. It linearly associated with mechanical and moisture permeability of the films. According to Xu \etal, in 2005, the film thickness was affected by many factors such as dry mass, drying conditions, alignment and distribution of components in the films \cite{r186}. With increase in glycerol concentration (acts as a plasticizer) also increases the film thickness \cite{r187}. In the development of the film 30\% glycerol is taken to enhance the overall functional quality of the film. Starch content from 3\% to 5\% increases the thickness of the film from 0.063 mm to 0.0140 mm \cite{r165}. For this reason 3\% corn starch was taken for the study. Acetic acid inside the corn-starch film acts as cross linker, which also increases the overall thickness of the starch film \cite{r189}. For this reason 5\% acetic acid solution was taken drop-wise just to balance the pH between 3 and 4. From the outcomes, it was observed (shown in Tab. \ref{tab4.6}) that nanocomposites starch films F1(as shown in Fig. \ref{fig4.14}), F2(as shown in Fig. \ref{fig4.15}) thickness increases from 0.019$\pm$0.33 mm to 0.019$\pm$0.90 mm with increase in crystallite size of the ZnO NPs and the starch film F3(as shown in Fig. \ref{fig4.16}) had the minimum obtained thickness value of 0.014$\pm$0.01 mm. ZnO NPs has been recognized safe by the US Food and Drug Administration (21CFR182.8991) \cite{r140}.

\item {\textbf{Moisture Content (MC):}}
Moisture holding capacity of the films was represented in terms of moisture content. It also affects the functional properties such as mechanical property and water vapor permeability, so it becomes an important film property. According to Mchugh \etal in 1994, stated that with increase in concentration of starch the MC also get enhanced \cite{r190}.Seligra \etal in 2014 \cite{r189}, said that citric acid improves the MC of the starch film. Ghanbarzadeh \etal in 2011 \cite{r127}, stated that citric acid entered between the starch polymer and thus decreases the interaction which ultimately results into enhancement of the MC. Glycerol or plasticizer content also increases the MC reported by Wang \etal in 2017. It was observed (shown in Tab. \ref{tab4.6}) that with addition of ZnO NPs, the MC also gets increases.

\item {\textbf{Swelling Index (SI):}}
M.A. Bertuzzi \etal (2007), stated that with increase in the concentration of the starch the swelling index also get enhanced \cite{r153}. Ghanbarzadeh \etal (2011) and McHugh \etal (1994), noted that with increase in concentration of the acetic acid and plasticizer in the corn-starch film alter the interaction of the two types of starch molecules (i.e. amylopectin and amylose) to the water molecules. It was observed (shown in Tab. \ref{tab4.6}) that chemically synthesized ZnO NPs embedded corn-starch film-F1 shows less SI than the green synthesized ZnO NPs embedded corn-starch film-F2 as 27 $\pm$0.35 and 28.57$\pm$0.14(SI\%), respectively. This may be due to the hydrophilic characteristics developed by the capping of the ZnO NPs by the neem extract.  The starch film-F3, developed with plasticizer and citric acid had maximum SI value i.e. 45.24$\pm$0.31(\%).

\begin{table*}[htb!]\centering
\caption{Functional properties of the films (Ref: Tab. \ref{tab4.5}).} \vskip 0.5cm
\begin{tabular}{|c|c|c|c|c|c|c|c|} \hline
Film&FT            &MC            &SI            &OP      &S             &WVPx10$^{-12}$& TS   \\
    &(mm)          &(\%)          &(\%)          &(\%)          &\%              &g$^{-1}$s$^{-1}$Pa$^{-1}$&(MPa)\\ \hline
F1&0.019$\pm$0.33&09.98$\pm$0.54&27.00$\pm$0.35&29.19$\pm$0.42&20.70$\pm$0.74&2.42$\pm$0.32&12.50$\pm$0.18\\ \hline
F2&0.019$\pm$0.90&10.13$\pm$0.78&28.57$\pm$0.14&30.43$\pm$0.28&21.86$\pm$0.42&2.85$\pm$0.06&11.34$\pm$0.27\\ \hline
F3&0.014$\pm$0.01&13.35$\pm$0.25&45.24$\pm$0.31&26.85$\pm$0.28&27.55$\pm$0.06&3.16$\pm$0.14&10.22$\pm$0.11\\ \hline
\end{tabular} \label{tab4.6}
\end{table*}
\item {\textbf{Opacity (OP):}}
Opacity is very important to study the enhancement of the shelf-life of the light sensitive food products. The film with high opacity reduces the transparency. According to Kim \etal in 2017, stated that the cross-linking in the starch occurred due to citric acid both crystalline and amorphous region which leading to decrease in the opacity \cite{r166}. Opacity increases with addition of ZnO NPs inside the starch-film. Comparison between the films i.e. F1, F2 and F3 (shown in Tab. \ref{tab4.6}), it was observed that F3 has the least opacity than the F2, F1 (where F1: embedded starch film with chemically synthesized ZnO NPs at 10 pH and F2: embedded starch film with green synthesized ZnO NPs at same 10 pH). Between F1 and F2 opacity of F2 is higher than the opacity of F1. Maryam Adilah  \etal (2017) shown the variation in the opacity of the film with increase the concentration of the mango kernel extracts (MKE) from 1 to 5\% \cite{r190}.

\item {\textbf{Solubility (S):}}
Solubility plays an important role in the proper selection of material for the food as it gives the information regarding the interaction of the films with water molecules. Maryam Adilah \etal (2017), studied that solubility of the starch films increased with the addition of hydrophilic mass in the films \cite{r190}. Seligra \etal (2016), examined that solubility of the cassava starch film decreases with addition of the citric acid \cite{r165}. It is observed that solubility decreases with addition of the ZnO NPs inside the starch-film. Comparison between the two films i.e. F1 and F2 (shown in Tab. \ref{tab4.6}), it was observed that F1 exhibits lesser solubility than the F2. This may be due to the hydrophilic characteristics developed by the capping of the ZnO NPs by the neem extract in F2. F3 showed the maximum solubility i.e.27.55$\pm$0.06(\%), due to high hydrophilic nature as compare to F1 and F2.

\item {\textbf{Water Vapor Permeability (WVP):}}
It is one of the main functional properties of packaging material. Water vapor permeability related to the water or moisture barrier properties of the packaging material. F. Mirjalili \etal ( 2017) stated that that WVP decreases with significantly addition of ZnO NPs \cite{r13} . It is observed that WVP decreases with addition of the ZnO NPs inside the starch-film. Comparison between two films i.e. F1 and F2 (shown in Tab. \ref{tab4.6}), it was observed that F2 showed slightly more WVP than the F1 i.e. 2.42$\pm$0.32 x 10$^{-12}$ and 2.85$\pm$0.50 x 10$^{-12}$ (g$^{-1}$s$^{-1}$Pa$^{-1}$), respectively. It was observed that solubility decreases with addition of the ZnO NPs inside the starch-film. F2 showed more WVP rate than F1, this may be due to the hydrophilic characteristics developed by the capping of the ZnO NPs by the neem extract in F2. SEM micrograph of the films also helped to explain the moisture migration mechanism in the films. Film-F3 showed the maximum WVP rate i.e. 3.16$\pm$0.14x 10$^{-12}$ (g$^{-1}$s$^{-1}$Pa$^{-1}$). Seligra \etal in 2016 explained that the cross-linking material inside the starch film also influence the WVP rate \cite{r165}.

\item {\textbf{Tensile Strength (TS):}}
It is again one of the most important functional properties of the packaging material. For the containment of the food product it is very important to analyze the tensile strength of the packaging material which also helps in transportation and distribution of food product. Merry Asria \etal (2015), stated that tensile strength increases with the addition of glycerol as plasticizer \cite{r187}. Seligra \etal (2016), stated that as the citric acid increases the tensile strength of the starch film also increases but at the same time elongation decreases \cite{r165}. The physical and mechanical stability also affected by the types and concentration of the plasticizers. Elastic limit is measured by modulus of elasticity due to which material elastic behavior is measured. In Glycerol (plasticizer polyalcohols), hydroxyl group interact with intermolecular and intramolecular through hydrogen bonds in the polymeric chain which provide flexibility to polymer due to which it attracted the consideration of researchers to develop bio polymer as the packaging material \cite{r133}.  It was observed (shown in Tab. \ref{tab4.6}) that TS increases with the addition of the ZnO NPs. Comparing films F1, F2 and F3; we observed that F1 showed the maximum TS whereas F3 showed the minimum i.e. 12.50$\pm$0.18and10.22$\pm$0.11 MPa (Mega Pascal), respectively. Due to the capping of neem extract over the ZnO NPs synthesized with green synthesis approach, film-F2 showed slightly lower TS i.e. 11.34$\pm$0.27 MPa, than film-F1 because of the higher hydrophilicity of the film-F2.

\item {\textbf{Biodegradability Analysis of the Synthesized Film:}}
Biodegradability of samples was checked using soil buried test. All specimens with dimension (1*1cm$^2$) of modified starch and composite were prepared from the films. Thereafter, plastic containers filled up to the surface with a mixture of uniform size soil and compost. Specimens were buried in the mixture below 2 cm from the mouth of the containers. Under controlled conditions (humidity 75$\pm$2\%, temperature 30$\pm$1$^\circ$C), weight loss of films were monitored after a fixed time interval of one week (Seligra \etal, 2016) \cite{r165}. It was observed (from Tab. \ref{tab4.7}) that film-F3 degraded first than F1 and F2. As it was also observed that biologically synthesized ZnO NPs embedded film i.e. F2 is degraded more quickly than F1.The methodology in testing biodegradability of the films was shown in Fig. 3.4. Babaee \etal (2015), also explained the degradation of the nanocomposites starch \cite{r191}.
\end{enumerate}
\begin{table}[htb!]\centering
\caption{Biological degradation of the different synthesized films.} \vskip 0.5cm
\begin{tabular}{|c|c|c|c|} \hline
Film                        & Initial weight & Final weight & Degradation \%           \\
(Ref:Tab. \ref{tab4.5})&                &              &                  \\ \hline
F1	                        &0.0345	         &0.0330         &04.350             \\ \hline
F2	                        &0.0359	         &0.0335	        &06.685             \\ \hline
F3	                        &0.0338	         &0.0311        &07.990            \\ \hline
\end{tabular} \label{tab4.7}
\end{table}

\subsection{Characterization of the ZnO NPs Embedded Starch Films}
\begin{enumerate}
\item {\textbf{XRD analysis of films:}}
The XRD of ZnO NPs (which were synthesized by two different approaches like biological approach in which green method is used and chemical approach in which sol-gel method is used) embedded corn-starch films were analyzed in Fig. \ref{fig4.17}. Shuzhen Ni \etal (2018), showed the similar results in miscibility of ZnO NPs inside the starch matrix \cite{r192}. In Figure 4.17(a) illustrated the XRD pattern of the biologically synthesized ZnO NPs embedded corn-starch film- F2 and the XRD pattern of the chemically synthesized ZnO NPs embedded-corn-starch film-F1 and (b) illustrated the comparison between the films F1, F2 and the pure ZnO NPs.
\begin{figure}[htb!]\centering
  \includegraphics[width=0.16\textwidth]{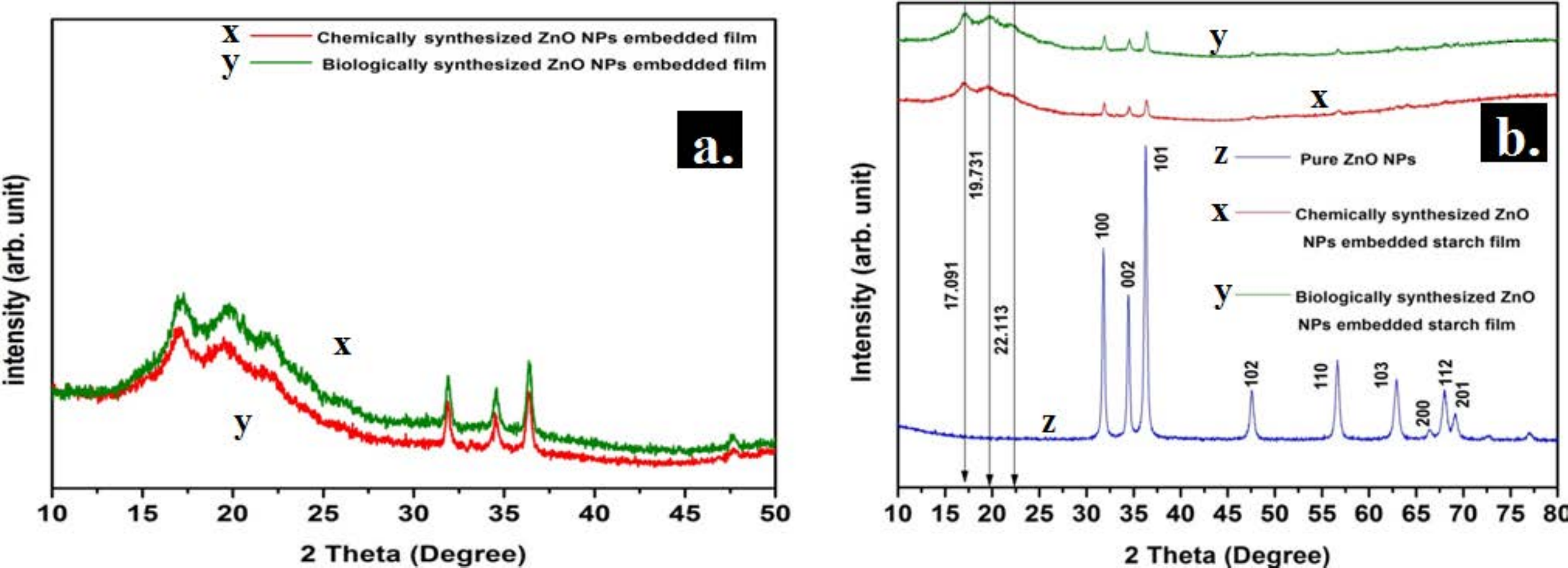}
  \caption{XRD pattern of (a) biochemically synthesized ZnO NPs embedded film (i.e. F2) and chemically synthesized ZnO NPs embedded film (i.e. F1) and (b) comparison between XRD pattern of film-F1, F2 and XRD pattern of pure ZnO NPs.}\label{fig4.17}
\end{figure}

Fig. \ref{fig4.17}(b) shows the crystallinity of the ZnO NPs and ZnO NPs embedded films (i.e. F1 and F2). The significant peaks appearing at 2$\theta$= 31.83$^\circ$, 34.46$^\circ$, 36.28$^\circ$, 47.68$^\circ$, 56.58$^\circ$, 62.98$^\circ$, 66.31$^\circ$, 68.13$^\circ$ and 69.23$^\circ$ correspond to the (100), (002), (101), (102), (110), (103), (200), (112) and (201) planes of the ZnO NPs crystal structure, respectively. Sharp peaks in XRD pattern shows high crystallinity without any presence of impurities as all the peaks present in XRD pattern matches with JCPDS card no. [01-070-2551]. In the corn-starch embedded film the characteristics peaks at 2$\theta$= 17.091$^\circ$, 19.731$^\circ$ and 22.113$^\circ$ almost completely disappeared as compare with the pure ZnO as shown in Fig. \ref{fig4.17}(b). The intermolecular interaction of between the ZnO NPs embedded inside the starch solution containing acetic acid and glycerol shows the excellent compatibility. The loss of the starch crystallinity is due to the gelatinization upon heating at 90-100$^\circ$C.

\item {\textbf{Scanning Electron Microscopy (SEM) Analysis of films:}} \index{Scanning Electron Microscopy (SEM) Analysis of films}
The micrographs of starch and nanocomposites films were studied by using SEM (as shown in Fig. \ref{fig4.18}). Fig. \ref{fig4.18} and Fig. \ref{fig4.19} at different resolution, revealed that the ZnO NPs embedded in the starch solution containing citric acid (as cross-linker) and glycerol (plasticizer) results into the uniform surface texture of the films with an even dispersion of ZnO NPs.
\begin{figure}[htb!]\centering
  \includegraphics[width=0.48\textwidth]{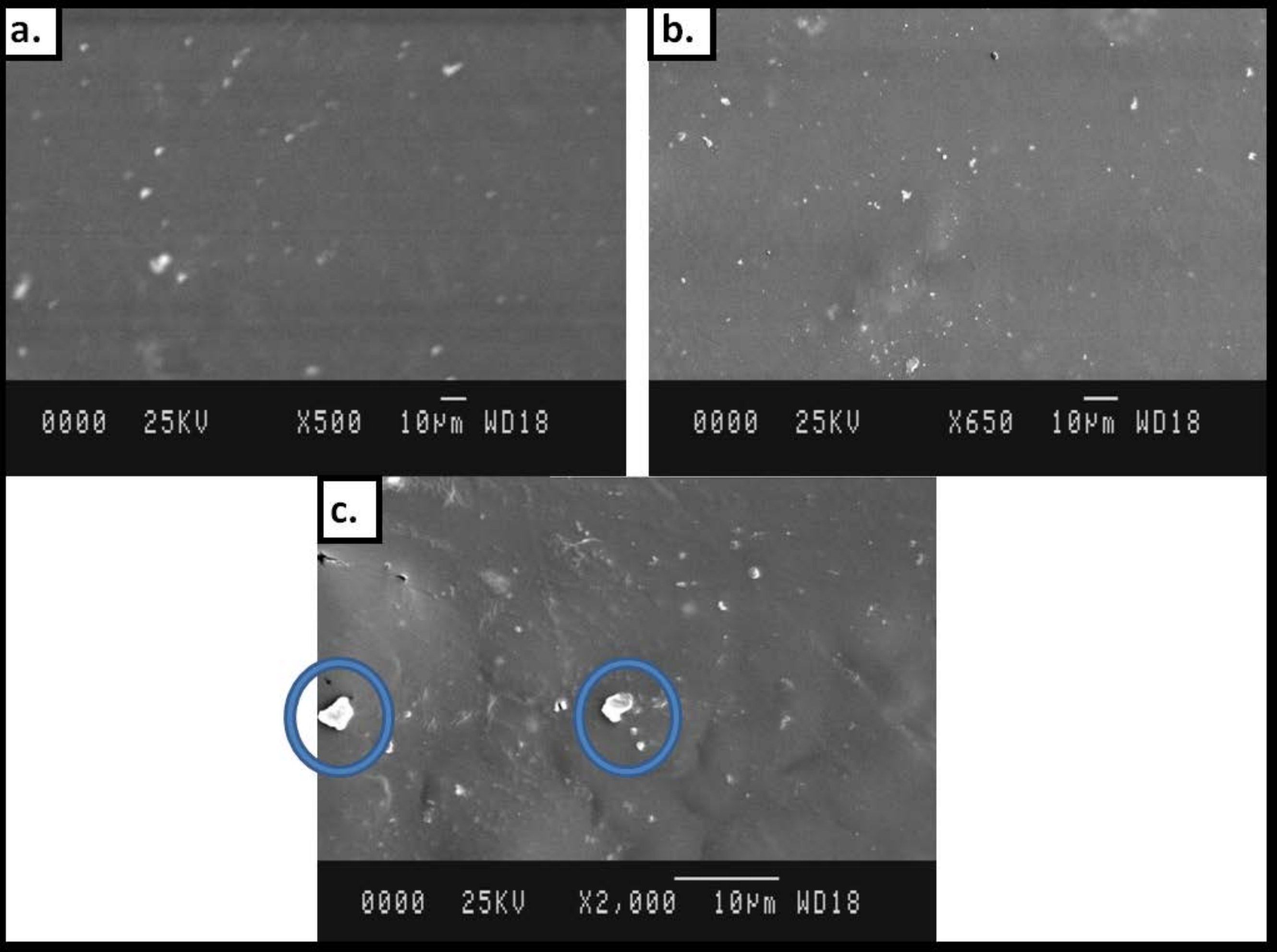}
  \caption{SEM images of biochemically synthesized ZnO NPs embedded corn starch-film; x500, (b) x650 and (c) x2,000 at 10 $\mu$m range.}\label{fig4.18}
\end{figure}
\begin{figure}[htb!]\centering
  \includegraphics[width=0.48\textwidth]{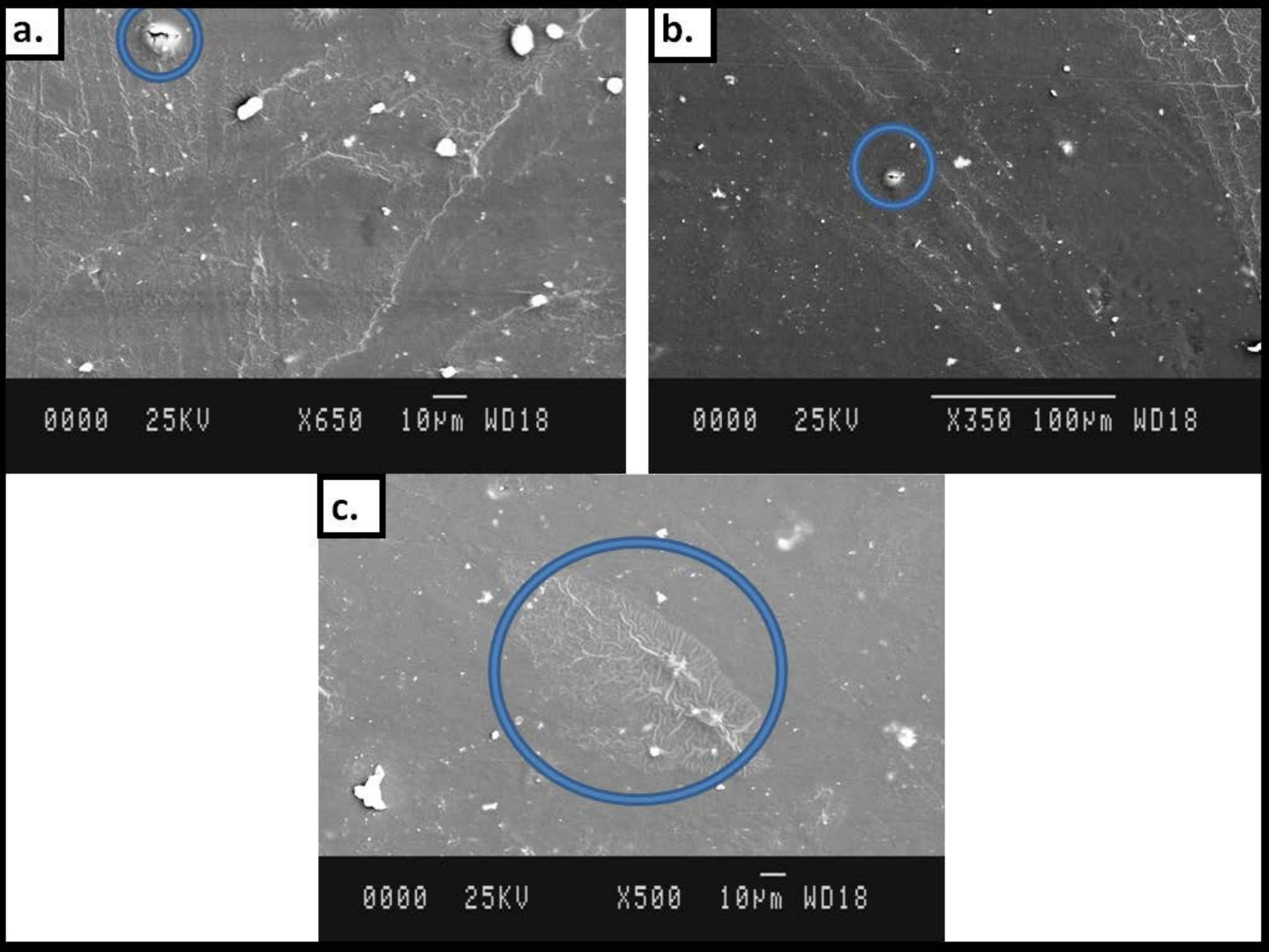}
  \caption{SEM images of chemically synthesized ZnO NPs embedded corn starch-film; (a) x650, (b) x350 and (c) x500 at 10 $\mu$m,100 $\mu$m and 10 $\mu$m range respectively.}\label{fig4.19}
\end{figure}
Small holes were observed on the surface of the embedded films [as shown in Fig. \ref{fig4.19} (a) and \ref{fig4.19}(b) through encircle]. This was probably due to the presence of agglomerate ZnO NPs which pierced the surface of the film. Some white dotted lumps were observed over the embedded films [as shown in Fig. \ref{fig4.18}(c) through encircle] this was probably due to the agglomeration of the ZnO NPs which results into the formation of the lumps. In Fig. \ref{fig4.19}(c) some winkle like patterns were seen which were shown trough encircle, which were probably due to the stretching of the films to experience its tensile strength manually.

Micrographs also showed that the film surface is homogeneous, smooth when ZnO NPs added inside the starch solution. Similar micrographs were discussed by Z. Luo \etal in their results in 2013 \cite{r193}. Through the results we concluded that the biochemically synthesized ZnO NPs embedded films were smoother than the chemically synthesized ZnO NPs embedded films.

\item {\textbf{Antibacterial Activity of the ZnO NPs Embedded Starch Films:}}
Antibacterial activity of ZnO embedded corn starch film against \emph{E. coli} (gram negative bacteria) and \emph{S. aureus} (gram positive bacteria) was studied using standard agar diffusion method. Antibacterial activity was performed against two bacterial strains as per guidelines (CLSI, M02-A12). These were grown overnight and diluted in Mueller-Hinton broth (MHB) to a cell density of 10$^5$ Colony Forming Unit (CFU)/mL. 100 $\mu$l of this culture was spread on the Mueller-Hinton Agar (MHA) plate and allowed to dry in a sterile condition.
\begin{figure}[htb!]\centering
  \includegraphics[width=0.48\textwidth]{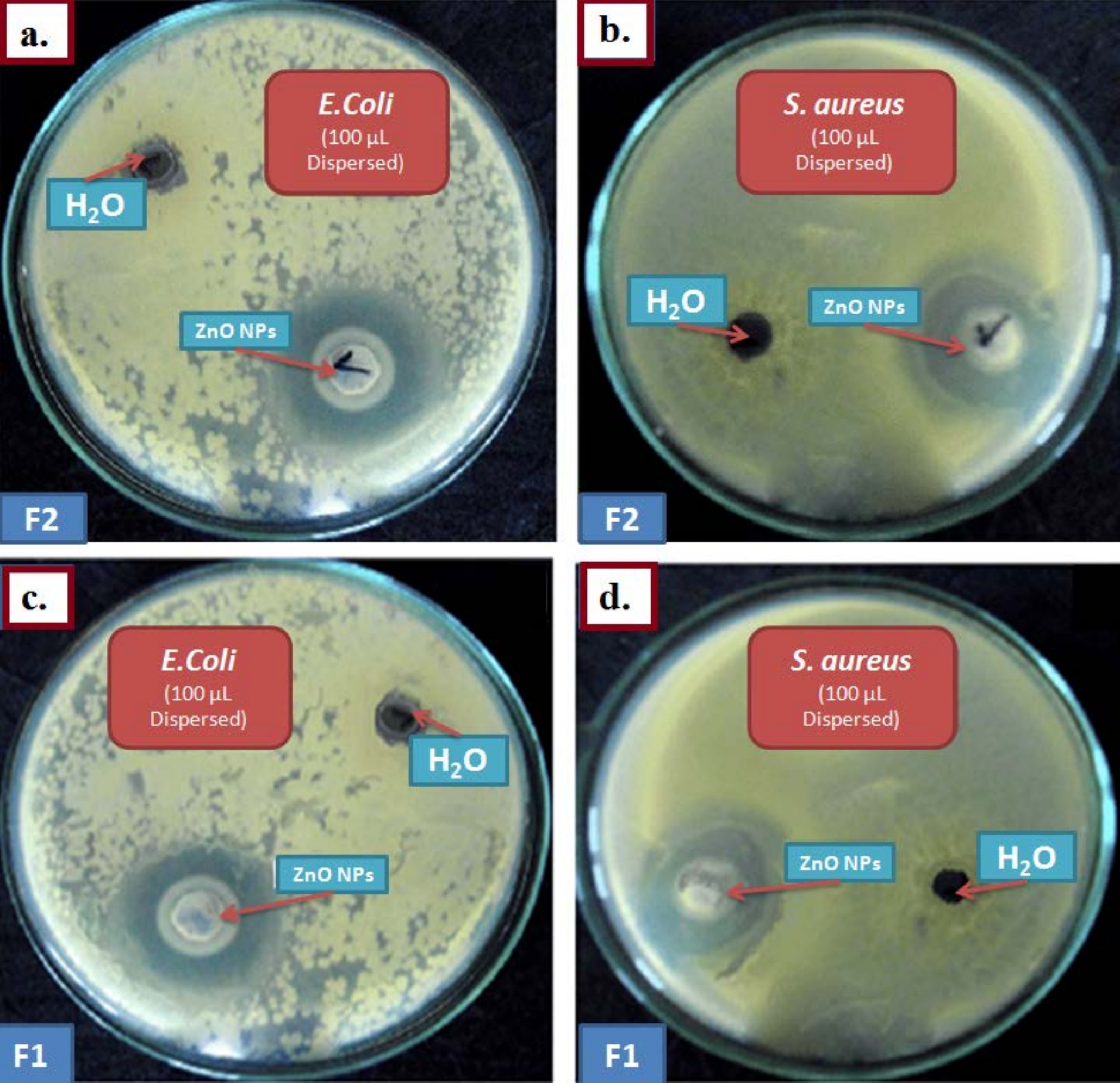}
  \caption{ZnO NPs embedded starch solution of film - F2 was added into the wells of two bacterial cultures; (a) \emph{E.coli} (gram negative bacteria) and (b) \emph{S.aureus} (gram positive bacteria) and ZnO NPs embedded starch solution of film-F1 was added into the wells of two same bacterial cultures shown in; (c) \emph{E.coli} and (d) \emph{S.aureus}}.
\label{fig4.20}
\end{figure}
For this starch solution which containing ZnO NPs (100 mg concentration) was added into well (6 mm width) on MHA plate. ZnO NPs embedded-starch solution was taken for the test because it is similar whether to test the film or to test the solution. The plate was incubated at 37$\pm$1$^\circ$C for 24 hrs. Antibacterial activity of ZnO NPs/starch solution was measured based on the zone of inhibition around the well infused with the rifampicin (RIF) and HPLC grade sterile water and biosynthesized silver nanoparticles/starch solution were used as positive and negative control respectively. The result of inhibition zone of Zn NPs embedded starch film was shown in Fig. \ref{fig4.20}. It was observed that ZnO embedded starch films showed high inhibitory action against \emph{E. coli} (gram negative bacteria) as compare to \emph{S.aureus} (gram positive bacteria).

\begin{table}[htb!]\centering
\caption{Zone of Inhibition of ZnO NPs embedded starch film solution.} \vskip 0.5cm
\begin{tabular}{|c|c|c|} \hline
Bacteria                        &NPs embedded film  &Inhibition zone(mm)       \\ \hline
\emph{E. coli}	        & F1 solution            & 12                         \\
	                            & F2 solution	         & 14	                      \\ \hline
\emph{S. aureus}	& F1 solution            & 5	                      \\
                                & F2 solution            & 6                          \\ \hline
\end{tabular} \label{tab4.8}
\end{table}
Film-F2 showed effective inhibitory as compare to the film-F1 in both cases as shown in the Tab. \ref{tab4.8} and Fig. \ref{fig4.20}. This may be due the presence of capping of neem extract around the green synthesized ZnO NPs, as neem itself has its own antimicrobial properties \cite{r140}. The zone of inhibition of ZnO NPs was also studied by F. Mirjalili \etal in 2017 \cite{r13}. Biochemically synthesized ZnO NPs (with neem) exhibits antimicrobial properties \cite{r139}. Jinxia Ma \etal in 2016, showed the antimicrobial effect of ZnO-starch nanocomposites and also explained its coating application \cite{r122}. S.A. Kadhum in 2017 explained the antibacterial effects of ZnO and SiO$_2$ NPs against gram positive and gram negative bacteria \cite{r35}. At the end the conclusion of the experiments were that ZnO NPs or ZnO NPs embedded starch films both showed antimicrobial effects against gram positive and gram negative bacteria.
\end{enumerate}

\subsection{Qualitative Analysis of the ZnO NPs Embedded Starch Film Wrapped over the Grapes}
Application of ZnO embedded nanocomposites corn-starch film (biosynthesized NPs) on grapes. Film-F2 is taken for the qualitative analysis of the grapes, as it exhibited higher inhibition zone (as shown in Tab. \ref{tab4.5}) against both gram positive and negative bacterial. F2 is synthesized with addition of biosynthesized ZnO NPs which also exhibits good functional properties Tab. \ref{tab4.6}. One set of Grapes were wrapped in Nano-composite film (WG) and other set was without film (NWG)
\begin{figure}[htb!]\centering
  \includegraphics[width=0.48\textwidth]{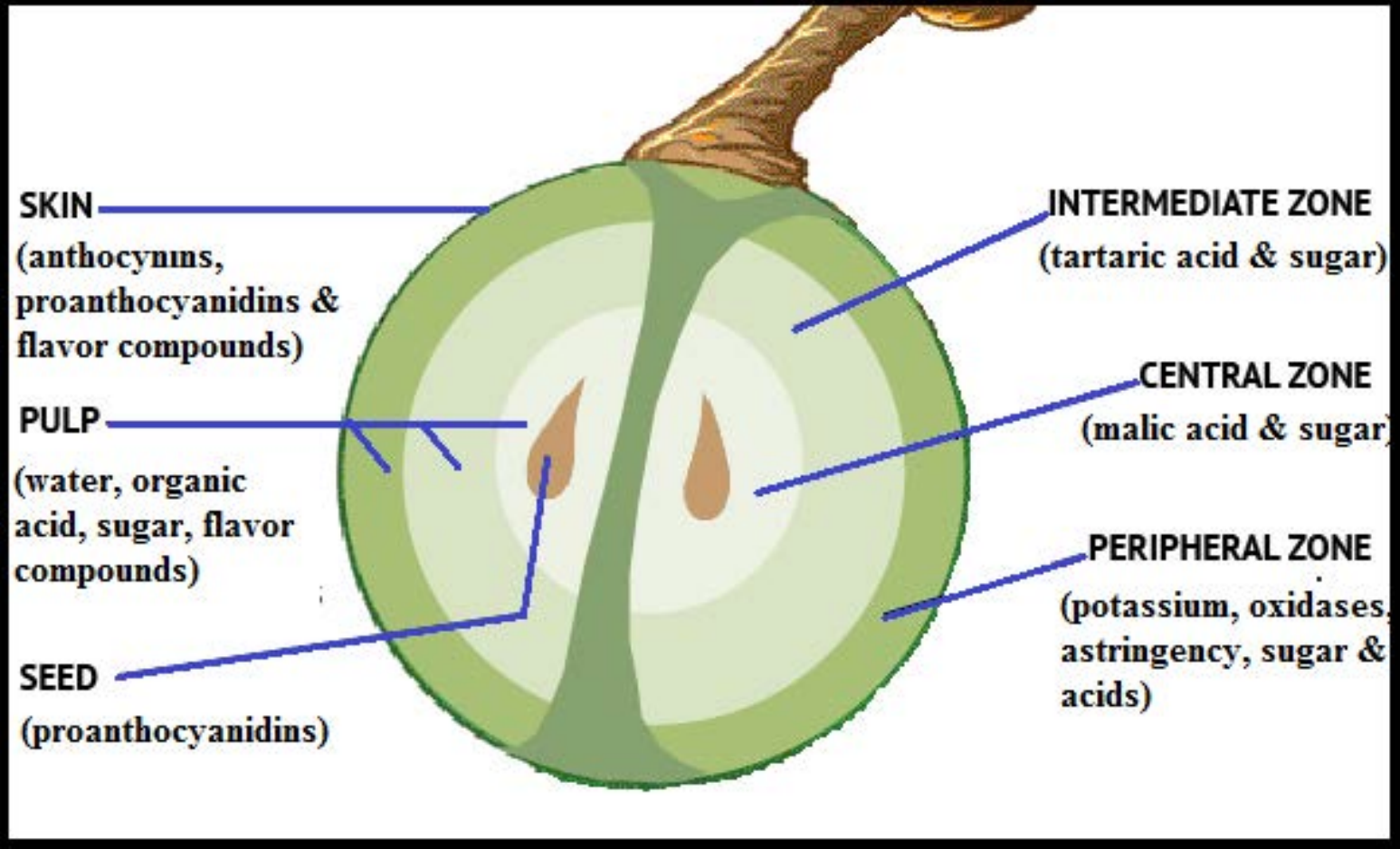}
  \caption{Well labeled diagram of the grapes components.}\label{fig4.21}
\end{figure}

The well labeled diagram of the grapes components is shown in Fig. \ref{fig4.21}. The physio-chemical properties of Grapes which were Weight of Grapes, pH, Color, Total Soluble Solids (TSS) and Acidity were studied for 7 days at room temperature (RT) $~$30$^\circ$C.
\begin{enumerate}
\item {\textbf{Weight, color, TSS, pH and Acidity Analysis of the Wrapped and Non-Wrapped Grapes:}}
It was observed (in Tab. \ref{tab4.9} and Tab. \ref{tab4.10}) that the wrapped grapes showed much better results as compare to the non-wrapped grapes. As the weight loss in wrapped grapes is pretty much slow as compare to the non-wrapped grapes. Acidity decreases with ripening of the grapes. The decreasing rate in the pH seems much higher in non-wrapped grapes as compare to the wrapped grapes. TSS in non-wrapped grapes at room temperature exhibited much higher compare to the wrapped grapes. Color effects were not seen in the wrapped grapes but there is drastically change in color of the non-wrapped grapes and this may be due to the loss of water or drying of the grapes. M.A Del Nobile \etal (2009), similar work was done, in which grapes were wrapped and observed for the period of 35 days \cite{r194}.
\begin{table}[htb!]\centering
\caption{Physio-chemical properties (TSS, pH and Acidity) of wrapped (W) and non-wrapped (NW) Grapes taken on a period of seven days at RT $~$30$^\circ$C.} \vskip 0.5cm
\begin{tabular}{|c|c|c|c|c|} \hline
Days & \multicolumn{2}{c|}{Weight(g)}   &\multicolumn{2}{c|}{Color}                         \\ \hline
     & W                  & NW                & W                    & NW     \\ \hline
	 &                    &                   & 38.57$\pm$0.003    & 37.32$\pm$0.005       \\
1     & 6.002$\pm$ & 5.678$\pm$  & -387.57$\pm$0.006  & -252.60$\pm$0.001     \\
     &    0.008                  &  0.001                 & 15.34$\pm$0.001   & 21.83$\pm$0.007      \\ \hline
    &                    &                   & 38.92$\pm$0.002    & 43.74$\pm$0.008       \\
2    & 5.365$\pm$   & 2.945$\pm$ & 01.57$\pm$0.006    & -02.08$\pm$0.002      \\
     &  0.003                  &  0.007                  & 19.33$\pm$0.007   & 18.55$\pm$0.006      \\ \hline
\end{tabular} \label{tab4.9}
\end{table}
\begin{table*}[htb!]\centering
\caption{Physio-chemical properties (Color and Weight) of wrapped (W) and non-wrapped (NW) Grapes taken on a 1$^{st}$ and 7$^{th}$ day at RT $~$30$^\circ$C.} \vskip 0.5cm
\begin{tabular}{|c|c|c|c|c|c|c|} \hline
Days&\multicolumn{2}{c|}{TSS}	    &\multicolumn{2}{c|}{pH}   &\multicolumn{2}{c|}{Acidity}      \\ \hline
	&W	            &NW	            &W	           &NW	           &W	          &NW             \\ \hline
1	&24.01$\pm$0.001&24.12$\pm$0.002&3.52$\pm$0.009& 3.71$\pm$0.004&3.75$\pm$0.006&3.81$\pm$0.003 \\ \hline
7	&25.94$\pm$0.005&29.03$\pm$0.004&3.36$\pm$0.005& 2.88$\pm$0.002&3.63$\pm$0.009&3.22$\pm$0.005 \\ \hline
\end{tabular} \label{tab4.10}
\end{table*}

\item {\textbf{Sensory Evaluation of the Wrapped and Non-Wrapped Grapes:}}
Sensory evaluation of coated and native samples of fruits was conducted using 9-point hedonic scale (as shown in Tab. \ref{tab4.11}) to check the quality and acceptability wrapped and non-wrapped grapes, after storage of 7 days by a panel of 5 semi-trained members. Stored samples qualities and acceptability were judged on the basis of visual aspect, color, texture, flavor, and overall acceptability.
\begin{table}[htb!]\centering
\caption{9-point hedonic scale for sensory evaluation.} \vskip 0.5cm
\begin{tabular}{|c|c|} \hline
\multicolumn{2}{|c|}{hedonic scale} \\ \hline
9&Like extremely            \\ \hline
8&Like very much            \\ \hline
7&Like moderately           \\ \hline
6&like slightly            \\ \hline
5&Neither like nor dislike  \\ \hline
4&Dislike slightly          \\ \hline
3&Dislike moderately        \\ \hline
2&Dislike very much         \\ \hline
1&Dislike extremely         \\ \hline
\end{tabular} \label{tab4.11}
\end{table}
\begin{table}[htb!]\centering
\caption{Points given by 5-semi trained panelist on the basis of 9-point hedonic scale for wrapped (W) and non-wrapped grapes.} \vskip 0.5cm
\begin{tabular}{|c|c|c|c|c|c|c|c|c|} \hline
S/No.       & \multicolumn{2}{c|}{Color} & \multicolumn{2}{c|}{Texture} & \multicolumn{2}{c|}{Flavor} &	 \multicolumn{2}{c|}{\small Overall} \\
& \multicolumn{2}{c|}{}        & \multicolumn{2}{c|}{}          & \multicolumn{2}{c|}{}        & \multicolumn{2}{c|}{\small acceptability}\\ \hline
Panel's     & W   & NW    &	W   & NW  & W  & NW &	W & NW        \\ \hline
1	        & 8   & 4	  & 7   & 3   & 8  & 6  &  8 & 5          \\ \hline
2	        & 6   & 5     & 8   & 2   & 9  & 7  & 8  & 4          \\ \hline
3	        & 7   & 3     & 5   & 4   & 7  & 5  & 7  & 4          \\ \hline
4	        & 8   & 4     & 6   & 4   & 8  & 5  & 9  & 3          \\ \hline
5	        & 7   & 6     & 7   & 2   & 7  & 4  & 7  & 4          \\ \hline
Average   	& 7.2 & 4.4   & 6.6	& 3   &	7.8& 5.4& 7.8&	4         \\ \hline
\end{tabular} \label{tab4.12}
\end{table}
On the basis of 9-point hedonic scale (as shown in Tab. \ref{tab4.11}) and the score card (as shown in Tab. \ref{tab4.12}) it was concluded that the wrapped grapes gives the better sensory results as compare to the non-wrapped grapes. The overall acceptability of wrapped grapes is much more than non-wrapped grapes with average score of 7.8 and 4, respectively.
\end{enumerate}
\section{Conclusion}
In the present study the ZnO nanoparticles embedded starch film was synthesized. Films biodegradability analysis, functional analysis, qualitative analysis and antimicrobial effectiveness were tested. ZnO nanoparticles were successfully synthesized chemically and alternatively by green synthesis approach as well. In chemical synthesis approach; sol-gel techniques was used and ZnO NPs were synthesized at different pH level of 7 to 11 pH. The confirmation of the successful synthesis of the ZnO NPs was characterized by XRD analysis which was further cross-checked and compared with the JCPDS database card no. [01-070-2551]. At 10 pH minimum crystallite size of ZnO nanoparticles with average crystallite size of 28nm was detected. It was observed that growth of ZnO nanoparticles is difficult at the lower concentration of OH$^-$ ions in the reaction mixture (i.e., pH$<$8) and as the concentration increases it leads to formation of hexagonal and spherical shaped nanostructures. UV-DRS of the chemically synthesized ZnO nanoparticles showed band gap of 3.11 to 3.19 eV and 3.12 to 3.25 eV before and after annealing respectively. FE-SEM result have shown the morphology of  ZnO nanoparticles which infers that ZnO nanoparticles synthesis at 10 pH produces pure and homogeneous nanoparticles with hexagonal and spherical structures unlike at 8 pH. In green synthesis approach; B1: 2\% Neem (Azadirachta indica) extract and B2: 3\% neem extract was used to synthesize the ZnO nanoparticles with the crystallite size of 36.34 nm and 32.68 nm respectively. Their crystallinity and morphology were determined by XRD and FESEM analysis respectively. Through, EDX the Zinc (Zn) and Oxygen (O) atomic ratios were observed with peaks; in which the ratios were 1.36:1 and 1.5:1 in chemically synthesized ZnO NPs at 8 and 10 pH respectively whereas in green synthesis approach the ratio was 1.22:1 at 10 pH. \emph{Azadirachta indica} (neem) extract was used in synthesis of ZnO nanoparticles due to its green, eco-friendly and inexpensive approach and it is easily available in India with reported medicinal and antimicrobial properties. UV-VIS value from 380 nm to 380.12 nm indicated the synthesis of the ZnO nanoparticles through green approach. Functional and characterization of synthesized starch films-F1, F2 and F3 (where, F1: embedded with chemically synthesized ZnO NPs at pH 10 is taken, F2: embedded with biologically synthesized ZnO NPs maintained at pH 10 and F3: starch film without embedded any ZnO nanoparticles) was done with the help of Film thickness (FT), moisture content (MC), swelling index (SI), opacity (OP), solubility (S), water vapor permeability (WVP), mechanical property (MP), XRD and SEM analysis. WVP increases form F1, F2 to F3 i.e. 2.42$\pm0.32\times10^{-12}$, 2.85$\pm0.50\times10^{-12}$ to 3.16$\pm0.14\times10^{-12}$ (gm$^{-1}$s$^{-1}$Pa$^{-1}$) respectively. It was observed that S, SI and MC increases from F1, F2 to F3. Thickness and opacity increases form F3, F1 to F1. Tensile strength increases form F3, F2 to F1 i.e. 12.50$\pm$0.18, 11.34$\pm$0.27 and 10.22$\pm$0.11 (MPa) respectively. According to 7 days biodegradability analysis it was observed that the biodegradation \% of films increases from F1, F2 to F3. In XRD of the ZnO NPs embedded corn-starch films the characteristics peaks at 2$\theta = 17.091^\circ$, 19.731$^\circ$ and 22.113$^\circ$ almost disappeared as compare with the pure ZnO peaks. Micrographs in SEM results showed that the film surface is homogeneous, smooth when ZnO NPs added inside the starch solution. Films solution of F2 showed greater antimicrobial effectiveness (inhibition zone) against the \emph{E.coli} (gran negative) and \emph{S.aureus} (gram positive) bacterial strains as compare to F1. In a qualitative analysis of the wrapped and non-wrapped grapes was observed by wrapping the grapes with film F1 and F2, in which it was observed that the film F2 wrapped grapes showed the least changes over the freshness of the grapes as compare to the film F1 wrapped grapes and the non-wrapped grapes; over a period of 7 days. The overall acceptance of the film F2 wrapped grapes is more as compare to the film F1 and non-wrapped grapes.

\section{Future Scope of Work}
In future different types of similar antimicrobial nanoparticles embedded starch films can be developed with the help of different types of metal nanoparticles and different types of starches. The full proof antibacterial mechanisms of the nanoparticles are still part of research. Many researchers concluded that the antibacterial behavior of the metal nanoparticles is due to the oxidative stress or ROS generation inside the bacterial cells. But the proper proof is still in consideration. More work should be done in biological synthesis of the nanoparticles with the help of \emph{Azadirachta indica} (Neem). Green synthesis approach is eco-friendly, economic and more beneficial as compare to the chemical synthesis process. So, more consideration should be there in biological synthesis process of the nanoparticles. Functional properties of the different types of starch films with better combination of corss-linking agents (such citric acid, ferulic acid, tannic acid etc.), plasticizers (such as sorbitol, glycerol etc.) and other stabilizing agents (such as proteins, whey proteins, pectins etc.) should be done. Films qualitative analysis should be done with different types of highly perishable food products. Toxicity test of different metal nanoparticles should be done.

\section*{Acknowledgments}
SAP (Special Assistance Program, UGC, Govt. of India) and TEQIP-III (MHRD, Govt. of India) grants for Panjab University Chandigarh (India) are duly acknowledged.


\section*{REFERENCES}



\bibliographystyle{model1a-num-names}%
\bibliography{RPSFerriteRev}

\begin{thebibliography}{53}
\expandafter\ifx\csname natexlab\endcsname\relax\def\natexlab#1{#1}\fi
\providecommand{\bibinfo}[2]{#2}
\ifx\xfnm\relax \def\xfnm[#1]{\unskip,\space#1}\fi
\bibitem[{Napierala and Stangierski(2007)}]{r1}
\bibinfo{author}{D.~M. Napierala}, \bibinfo{author}{J.~Stangierski},
  \bibinfo{journal}{Acta Agrophysica} \bibinfo{volume}{9}
  (\bibinfo{year}{2007}) \bibinfo{pages}{123--133}.
\bibitem[{Liu(2006)}]{r2}
\bibinfo{author}{L.~Liu}, \bibinfo{journal}{San Jose State University Packaging
  Engineering} \bibinfo{volume}{13} (\bibinfo{year}{2006})
  \bibinfo{pages}{1348--1368}.
\bibitem[{Rands et~al.(2010)Rands, Adams, Bennun, Butchart, Clements, Coomes,
  Entwistle, Hodge, Kapos, Scharlemann et~al.}]{r3}
\bibinfo{author}{M.~R. Rands}, \bibinfo{author}{W.~M. Adams},
  \bibinfo{author}{L.~Bennun}, \bibinfo{author}{S.~H. Butchart},
  \bibinfo{author}{A.~Clements}, \bibinfo{author}{D.~Coomes},
  \bibinfo{author}{A.~Entwistle}, \bibinfo{author}{I.~Hodge},
  \bibinfo{author}{V.~Kapos}, \bibinfo{author}{J.~P. Scharlemann}, et~al.,
  \bibinfo{journal}{science} \bibinfo{volume}{329} (\bibinfo{year}{2010})
  \bibinfo{pages}{1298--1303}.
\bibitem[{Espitia et~al.(2014)Espitia, Du, de~Jes{\'u}s Avena-Bustillos,
  Soares, and McHugh}]{r4}
\bibinfo{author}{P.~J.~P. Espitia}, \bibinfo{author}{W.-X. Du},
  \bibinfo{author}{R.~de~Jes{\'u}s Avena-Bustillos}, \bibinfo{author}{N.~d.
  F.~F. Soares}, \bibinfo{author}{T.~H. McHugh}, \bibinfo{journal}{Food
  hydrocolloids} \bibinfo{volume}{35} (\bibinfo{year}{2014})
  \bibinfo{pages}{287--296}.
\bibitem[{Jim{\'e}nez et~al.(2012)Jim{\'e}nez, Fabra, Talens, and Chiralt}]{r5}
\bibinfo{author}{A.~Jim{\'e}nez}, \bibinfo{author}{M.~J. Fabra},
  \bibinfo{author}{P.~Talens}, \bibinfo{author}{A.~Chiralt},
  \bibinfo{journal}{Food and Bioprocess Technology} \bibinfo{volume}{5}
  (\bibinfo{year}{2012}) \bibinfo{pages}{2058--2076}.
\bibitem[{Caz{\'o}n et~al.(2017)Caz{\'o}n, Velazquez, Ram{i}rez, and
  V{\'a}zquez}]{r6}
\bibinfo{author}{P.~Caz{\'o}n}, \bibinfo{author}{G.~Velazquez},
  \bibinfo{author}{J.~A. Ram{i}rez}, \bibinfo{author}{M.~V{\'a}zquez},
  \bibinfo{journal}{Food Hydrocolloids} \bibinfo{volume}{68}
  (\bibinfo{year}{2017}) \bibinfo{pages}{136--148}.
\bibitem[{Sharlina et~al.(2017)Sharlina, Yaacob, Lazim, Fazry, Lim, Abdullah,
  Noordin, and Kumaran}]{r7}
\bibinfo{author}{M.~E. Sharlina}, \bibinfo{author}{W.~Yaacob},
  \bibinfo{author}{A.~M. Lazim}, \bibinfo{author}{S.~Fazry},
  \bibinfo{author}{S.~J. Lim}, \bibinfo{author}{S.~Abdullah},
  \bibinfo{author}{A.~Noordin}, \bibinfo{author}{M.~Kumaran},
  \bibinfo{journal}{Food chemistry} \bibinfo{volume}{220}
  (\bibinfo{year}{2017}) \bibinfo{pages}{225--232}.
\bibitem[{Nawab et~al.(2017)Nawab, Alam, Haq, Lutfi, and Hasnain}]{r8}
\bibinfo{author}{A.~Nawab}, \bibinfo{author}{F.~Alam}, \bibinfo{author}{M.~A.
  Haq}, \bibinfo{author}{Z.~Lutfi}, \bibinfo{author}{A.~Hasnain},
  \bibinfo{journal}{International journal of biological macromolecules}
  \bibinfo{volume}{98} (\bibinfo{year}{2017}) \bibinfo{pages}{869--876}.
\bibitem[{Sorrentino et~al.(2007)Sorrentino, Gorrasi, and Vittoria}]{r9}
\bibinfo{author}{A.~Sorrentino}, \bibinfo{author}{G.~Gorrasi},
  \bibinfo{author}{V.~Vittoria}, \bibinfo{journal}{Trends in Food Science \&
  Technology} \bibinfo{volume}{18} (\bibinfo{year}{2007})
  \bibinfo{pages}{84--95}.
\bibitem[{Noorbakhsh-Soltani et~al.(2018)Noorbakhsh-Soltani, Zerafat, and
  Sabbaghi}]{r10}
\bibinfo{author}{S.~Noorbakhsh-Soltani}, \bibinfo{author}{M.~Zerafat},
  \bibinfo{author}{S.~Sabbaghi}, \bibinfo{journal}{Carbohydrate polymers}
  \bibinfo{volume}{189} (\bibinfo{year}{2018}) \bibinfo{pages}{48--55}.
\bibitem[{Brody et~al.(2008)Brody, Bugusu, Han, Sand, and Mchugh}]{r11}
\bibinfo{author}{A.~L. Brody}, \bibinfo{author}{B.~Bugusu},
  \bibinfo{author}{J.~H. Han}, \bibinfo{author}{C.~K. Sand},
  \bibinfo{author}{T.~H. Mchugh}, \bibinfo{journal}{Journal of food science}
  \bibinfo{volume}{73} (\bibinfo{year}{2008}) \bibinfo{pages}{107--116}.
\bibitem[{Pal(2017)}]{r12}
\bibinfo{author}{M.~Pal}, \bibinfo{journal}{J Food e Microbiol Safety Hyg}
  \bibinfo{volume}{2} (\bibinfo{year}{2017}) \bibinfo{pages}{121}.
\bibitem[{Mirjalili and Yassini~Ardekani(2017)}]{r13}
\bibinfo{author}{F.~Mirjalili}, \bibinfo{author}{A.~Yassini~Ardekani},
  \bibinfo{journal}{Journal of food process engineering} \bibinfo{volume}{40}
  (\bibinfo{year}{2017}) \bibinfo{pages}{e12561}.
\bibitem[{Rana and Kalaichelvan(2011)}]{r14}
\bibinfo{author}{S.~Rana}, \bibinfo{author}{P.~Kalaichelvan},
  \bibinfo{journal}{Antibacterial Activities of Metal Nanoparticles}
  \bibinfo{volume}{11} (\bibinfo{year}{2011}) \bibinfo{pages}{21--23}.
\bibitem[{Ghavidel and Prakash(2007)}]{r19}
\bibinfo{author}{R.~A. Ghavidel}, \bibinfo{author}{J.~Prakash},
  \bibinfo{journal}{LWT-Food Science and Technology} \bibinfo{volume}{40}
  (\bibinfo{year}{2007}) \bibinfo{pages}{1292--1299}.
\bibitem[{Ghasemlou et~al.(2011)Ghasemlou, Khodaiyan, and Oromiehie}]{r20}
\bibinfo{author}{M.~Ghasemlou}, \bibinfo{author}{F.~Khodaiyan},
  \bibinfo{author}{A.~Oromiehie}, \bibinfo{journal}{Carbohydrate Polymers}
  \bibinfo{volume}{84} (\bibinfo{year}{2011}) \bibinfo{pages}{477--483}.
\bibitem[{Fai et~al.(2016)Fai, de~Souza, de~Barros, Bruno, Ferreira, and
  de~Andrade~Gon{c}alves}]{r21}
\bibinfo{author}{A.~E.~C. Fai}, \bibinfo{author}{M.~R.~A. de~Souza},
  \bibinfo{author}{S.~T. de~Barros}, \bibinfo{author}{N.~V. Bruno},
  \bibinfo{author}{M.~S.~L. Ferreira}, \bibinfo{author}{{\'E}.~C.~B.
  de~Andrade~Gon{c}alves}, \bibinfo{journal}{Postharvest Biology and
  Technology} \bibinfo{volume}{112} (\bibinfo{year}{2016})
  \bibinfo{pages}{194--204}.
\bibitem[{Chiumarelli and Hubinger(2012)}]{r179}
\bibinfo{author}{M.~Chiumarelli}, \bibinfo{author}{M.~D. Hubinger},
  \bibinfo{journal}{Food hydrocolloids} \bibinfo{volume}{28}
  (\bibinfo{year}{2012}) \bibinfo{pages}{59--67}.
\bibitem[{Cao et~al.(2007)Cao, Fu, and He}]{r130}
\bibinfo{author}{N.~Cao}, \bibinfo{author}{Y.~Fu}, \bibinfo{author}{J.~He},
  \bibinfo{journal}{Food Hydrocolloids} \bibinfo{volume}{21}
  (\bibinfo{year}{2007}) \bibinfo{pages}{575--584}.
\bibitem[{Fakhouri et~al.(2015)Fakhouri, Martelli, Caon, Velasco, and
  Mei}]{r116}
\bibinfo{author}{F.~M. Fakhouri}, \bibinfo{author}{S.~M. Martelli},
  \bibinfo{author}{T.~Caon}, \bibinfo{author}{J.~I. Velasco},
  \bibinfo{author}{L.~H.~I. Mei}, \bibinfo{journal}{Postharvest Biology and
  Technology} \bibinfo{volume}{109} (\bibinfo{year}{2015})
  \bibinfo{pages}{57--64}.
\bibitem[{Nafchi et~al.(2012)Nafchi, Alias, Mahmud, and Robal}]{r119}
\bibinfo{author}{A.~M. Nafchi}, \bibinfo{author}{A.~K. Alias},
  \bibinfo{author}{S.~Mahmud}, \bibinfo{author}{M.~Robal},
  \bibinfo{journal}{Journal of food engineering} \bibinfo{volume}{113}
  (\bibinfo{year}{2012}) \bibinfo{pages}{511--519}.
\bibitem[{Adjouman et~al.(2017)Adjouman, Nindjin, Tetchi, Dalcq, Amani, and
  Sindic}]{r168}
\bibinfo{author}{Y.~D. Adjouman}, \bibinfo{author}{C.~Nindjin},
  \bibinfo{author}{F.~A. Tetchi}, \bibinfo{author}{A.-C. Dalcq},
  \bibinfo{author}{N.~G. Amani}, \bibinfo{author}{M.~Sindic},
  \bibinfo{journal}{Journal of Food Processing and Technology}
  \bibinfo{volume}{8} (\bibinfo{year}{2017}) \bibinfo{pages}{665}.
\bibitem[{Babaee et~al.(2015)Babaee, Jonoobi, Hamzeh, and Ashori}]{r191}
\bibinfo{author}{M.~Babaee}, \bibinfo{author}{M.~Jonoobi},
  \bibinfo{author}{Y.~Hamzeh}, \bibinfo{author}{A.~Ashori},
  \bibinfo{journal}{Carbohydrate polymers} \bibinfo{volume}{132}
  (\bibinfo{year}{2015}) \bibinfo{pages}{1--8}.
\bibitem[{Colivet and Carvalho(2017)}]{r171}
\bibinfo{author}{J.~Colivet}, \bibinfo{author}{R.~Carvalho},
  \bibinfo{journal}{Industrial crops and products} \bibinfo{volume}{95}
  (\bibinfo{year}{2017}) \bibinfo{pages}{599--607}.
\bibitem[{Warren(1990)}]{r180}
\bibinfo{author}{B.~E. Warren}, \bibinfo{title}{X-ray Diffraction},
  \bibinfo{publisher}{Courier Corporation}, \bibinfo{year}{1990}.
\bibitem[{Mote et~al.(2012)Mote, Purushotham, and Dole}]{r175}
\bibinfo{author}{V.~Mote}, \bibinfo{author}{Y.~Purushotham},
  \bibinfo{author}{B.~Dole}, \bibinfo{journal}{Journal of Theoretical and
  Applied Physics} \bibinfo{volume}{6} (\bibinfo{year}{2012})
  \bibinfo{pages}{6}.
\bibitem[{Xu and Wang(2011)}]{r183}
\bibinfo{author}{S.~Xu}, \bibinfo{author}{Z.~L. Wang}, \bibinfo{journal}{Nano
  Research} \bibinfo{volume}{4} (\bibinfo{year}{2011})
  \bibinfo{pages}{1013--1098}.
\bibitem[{Asria et~al.(2015)Asria, Elizarni, and Samah}]{r186}
\bibinfo{author}{M.~Asria}, \bibinfo{author}{Elizarni},
  \bibinfo{author}{d.~S.~D. Samah}, in: \bibinfo{booktitle}{AIP Conference
  Proceedings}, volume \bibinfo{volume}{1699(1)}, \bibinfo{organization}{AIP
  Publishing}, p. \bibinfo{pages}{040011}.
\bibitem[{Wawro and Kazimierczak(2008)}]{r187}
\bibinfo{author}{D.~Wawro}, \bibinfo{author}{J.~Kazimierczak},
  \bibinfo{journal}{Fibres \& Textiles in Eastern Europe} \bibinfo{volume}{16}
  (\bibinfo{year}{2008}) \bibinfo{pages}{71}.
\bibitem[{Alias et~al.(2010)Alias, Ismail, and Mohamad}]{r60}
\bibinfo{author}{S.~Alias}, \bibinfo{author}{A.~Ismail},
  \bibinfo{author}{A.~Mohamad}, \bibinfo{journal}{Journal of Alloys and
  Compounds} \bibinfo{volume}{499} (\bibinfo{year}{2010})
  \bibinfo{pages}{231--237}.
\bibitem[{Burrell(2003)}]{r15}
\bibinfo{author}{M.~Burrell}, \bibinfo{journal}{J.\ Experimental Botany}
  \bibinfo{volume}{54} (\bibinfo{year}{2003}).
\bibitem[{Bhuyan et~al.(2015)Bhuyan, Mishra, Khanuja, Prasad, and Varma}]{r140}
\bibinfo{author}{T.~Bhuyan}, \bibinfo{author}{K.~Mishra},
  \bibinfo{author}{M.~Khanuja}, \bibinfo{author}{R.~Prasad},
  \bibinfo{author}{A.~Varma}, \bibinfo{journal}{Materials Science in
  Semiconductor Processing} \bibinfo{volume}{32} (\bibinfo{year}{2015})
  \bibinfo{pages}{55--61}.
\bibitem[{Asif(2012)}]{r141}
\bibinfo{author}{M.~Asif}, \bibinfo{journal}{Journal of Pharmacognosy and
  phytochemistry} \bibinfo{volume}{1} (\bibinfo{year}{2012})
  \bibinfo{pages}{78--83}.
\bibitem[{Kanmani and Rhim(2014)}]{r121}
\bibinfo{author}{P.~Kanmani}, \bibinfo{author}{J.-W. Rhim},
  \bibinfo{journal}{Carbohydrate Polymers} \bibinfo{volume}{106}
  (\bibinfo{year}{2014}) \bibinfo{pages}{190--199}.
\bibitem[{Zhou et~al.(2017)Zhou, Fang, Gong, Xiao, Xie, Liu, and Cao}]{r37}
\bibinfo{author}{Y.~Zhou}, \bibinfo{author}{X.~Fang},
  \bibinfo{author}{Y.~Gong}, \bibinfo{author}{A.~Xiao},
  \bibinfo{author}{Y.~Xie}, \bibinfo{author}{L.~Liu}, \bibinfo{author}{Y.~Cao},
  \bibinfo{journal}{Nanomaterials} \bibinfo{volume}{7} (\bibinfo{year}{2017})
  \bibinfo{pages}{91}.
\bibitem[{Lennern{\"a}s and Abrahamsson(2005)}]{r34}
\bibinfo{author}{H.~Lennern{\"a}s}, \bibinfo{author}{B.~Abrahamsson},
  \bibinfo{journal}{Journal of pharmacy and pharmacology} \bibinfo{volume}{57}
  (\bibinfo{year}{2005}) \bibinfo{pages}{273--285}.
\bibitem[{Azizi et~al.(2014)Azizi, Ahmad, Namvar, and Mohamad}]{r137}
\bibinfo{author}{S.~Azizi}, \bibinfo{author}{M.~B. Ahmad},
  \bibinfo{author}{F.~Namvar}, \bibinfo{author}{R.~Mohamad},
  \bibinfo{journal}{Materials Letters} \bibinfo{volume}{116}
  (\bibinfo{year}{2014}) \bibinfo{pages}{275--277}.
\bibitem[{Santhoshkumar et~al.(2017)Santhoshkumar, Kumar, and
  Rajeshkumar}]{r195}
\bibinfo{author}{J.~Santhoshkumar}, \bibinfo{author}{S.~V. Kumar},
  \bibinfo{author}{S.~Rajeshkumar}, \bibinfo{journal}{Resource-Efficient
  Technologies} \bibinfo{volume}{3} (\bibinfo{year}{2017})
  \bibinfo{pages}{459--465}.
\bibitem[{Arvanitoyannis et~al.(1998)Arvanitoyannis, Nakayama, and Aiba}]{r145}
\bibinfo{author}{I.~Arvanitoyannis}, \bibinfo{author}{A.~Nakayama},
  \bibinfo{author}{S.-i. Aiba}, \bibinfo{journal}{Carbohydrate Polymers}
  \bibinfo{volume}{36} (\bibinfo{year}{1998}) \bibinfo{pages}{105--119}.
\bibitem[{Elumalai and Velmurugan(2015)}]{r196}
\bibinfo{author}{K.~Elumalai}, \bibinfo{author}{S.~Velmurugan},
  \bibinfo{journal}{Applied Surface Science} \bibinfo{volume}{345}
  (\bibinfo{year}{2015}) \bibinfo{pages}{329--336}.
\bibitem[{Seligra et~al.(2016)Seligra, Jaramillo, Fam{\'a}, and Goyanes}]{r165}
\bibinfo{author}{P.~G. Seligra}, \bibinfo{author}{C.~M. Jaramillo},
  \bibinfo{author}{L.~Fam{\'a}}, \bibinfo{author}{S.~Goyanes},
  \bibinfo{journal}{Carbohydrate polymers} \bibinfo{volume}{138}
  (\bibinfo{year}{2016}) \bibinfo{pages}{66--74}.
\bibitem[{McHugh and Krochta(1994)}]{r189}
\bibinfo{author}{T.~H. McHugh}, \bibinfo{author}{J.~M. Krochta},
  \bibinfo{journal}{Journal of agricultural and food chemistry}
  \bibinfo{volume}{42} (\bibinfo{year}{1994}) \bibinfo{pages}{841--845}.
\bibitem[{Adilah et~al.(2018)Adilah, Jamilah, and Hanani}]{r190}
\bibinfo{author}{Z.~M. Adilah}, \bibinfo{author}{B.~Jamilah},
  \bibinfo{author}{Z.~N. Hanani}, \bibinfo{journal}{Food hydrocolloids}
  \bibinfo{volume}{74} (\bibinfo{year}{2018}) \bibinfo{pages}{207--218}.
\bibitem[{Ren et~al.(2018)Ren, Dang, Pollet, and Av{\'e}rous}]{r127}
\bibinfo{author}{J.~Ren}, \bibinfo{author}{K.~Dang},
  \bibinfo{author}{E.~Pollet}, \bibinfo{author}{L.~Av{\'e}rous},
  \bibinfo{journal}{Polymers} \bibinfo{volume}{10} (\bibinfo{year}{2018})
  \bibinfo{pages}{808}.
\bibitem[{Bergo et~al.(2009)Bergo, Sobral, and Prison}]{r153}
\bibinfo{author}{P.~Bergo}, \bibinfo{author}{P.~Sobral},
  \bibinfo{author}{J.~Prison}, \bibinfo{title}{Physical properties of cassava
  starch films containing glycerol}, \bibinfo{year}{2009}.
\bibitem[{Kim et~al.(2017)Kim, Jane, and Lamsal}]{r166}
\bibinfo{author}{H.-Y. Kim}, \bibinfo{author}{J.-l. Jane},
  \bibinfo{author}{B.~Lamsal}, \bibinfo{journal}{Industrial crops and products}
  \bibinfo{volume}{95} (\bibinfo{year}{2017}) \bibinfo{pages}{175--183}.
\bibitem[{Menzel et~al.(2013)Menzel, Olsson, Plivelic, Andersson, Johansson,
  Kuktaite, J{\"a}rnstr{\"o}m, and Koch}]{r133}
\bibinfo{author}{C.~Menzel}, \bibinfo{author}{E.~Olsson},
  \bibinfo{author}{T.~S. Plivelic}, \bibinfo{author}{R.~Andersson},
  \bibinfo{author}{C.~Johansson}, \bibinfo{author}{R.~Kuktaite},
  \bibinfo{author}{L.~J{\"a}rnstr{\"o}m}, \bibinfo{author}{K.~Koch},
  \bibinfo{journal}{Carbohydrate polymers} \bibinfo{volume}{96}
  (\bibinfo{year}{2013}) \bibinfo{pages}{270--276}.
\bibitem[{Ni et~al.(2018)Ni, Zhang, Dai, and Xiao}]{r192}
\bibinfo{author}{S.~Ni}, \bibinfo{author}{H.~Zhang}, \bibinfo{author}{H.~Dai},
  \bibinfo{author}{H.~Xiao}, \bibinfo{journal}{Polymers} \bibinfo{volume}{10}
  (\bibinfo{year}{2018}) \bibinfo{pages}{1260}.
\bibitem[{Luo et~al.(2013)Luo, Cheng, Chen, Fu, Peng, Luo, and Nie}]{r193}
\bibinfo{author}{Z.~Luo}, \bibinfo{author}{W.~Cheng},
  \bibinfo{author}{H.~Chen}, \bibinfo{author}{X.~Fu},
  \bibinfo{author}{X.~Peng}, \bibinfo{author}{F.~Luo},
  \bibinfo{author}{L.~Nie}, \bibinfo{journal}{Journal of agricultural and food
  chemistry} \bibinfo{volume}{61} (\bibinfo{year}{2013})
  \bibinfo{pages}{4631--4638}.
\bibitem[{Sharma et~al.(2009)Sharma, Yngard, and Lin}]{r139}
\bibinfo{author}{V.~K. Sharma}, \bibinfo{author}{R.~A. Yngard},
  \bibinfo{author}{Y.~Lin}, \bibinfo{journal}{Advances in colloid and interface
  science} \bibinfo{volume}{145} (\bibinfo{year}{2009})
  \bibinfo{pages}{83--96}.
\bibitem[{Ma et~al.(2016)Ma, Zhu, Tian, and Wang}]{r122}
\bibinfo{author}{J.~Ma}, \bibinfo{author}{W.~Zhu}, \bibinfo{author}{Y.~Tian},
  \bibinfo{author}{Z.~Wang}, \bibinfo{journal}{Nanoscale research letters}
  \bibinfo{volume}{11} (\bibinfo{year}{2016}) \bibinfo{pages}{200}.
\bibitem[{Kadhum(2017)}]{r35}
\bibinfo{author}{S.~A. Kadhum}, \bibinfo{journal}{Biomedical \& Pharmacology
  Journal} \bibinfo{volume}{10} (\bibinfo{year}{2017}) \bibinfo{pages}{1701}.
\bibitem[{Del~Nobile et~al.(2009)Del~Nobile, Conte, Scrocco, Brescia, Speranza,
  Sinigaglia, Perniola, and Antonacci}]{r194}
\bibinfo{author}{M.~A. Del~Nobile}, \bibinfo{author}{A.~Conte},
  \bibinfo{author}{C.~Scrocco}, \bibinfo{author}{I.~Brescia},
  \bibinfo{author}{B.~Speranza}, \bibinfo{author}{M.~Sinigaglia},
  \bibinfo{author}{R.~Perniola}, \bibinfo{author}{D.~Antonacci},
  \bibinfo{journal}{Postharvest Biology and Technology} \bibinfo{volume}{51}
  (\bibinfo{year}{2009}) \bibinfo{pages}{21--26}.

\end{thebibliography}



%


\end{document}